\documentclass[
aps,%
10pt,%
onecolumn,%
nobibnotes,%
nofootinbib,
superscriptaddress,%
noshowpacs,%
centertags]%
{revtex4}

\usepackage{graphicx,amssymb,color}

\begin{document}

\title{Non-linear diffusive shock acceleration: A recipe for injection of electrons}

\author{\firstname{Bojan} \surname{Arbutina}}
\email[]{arbo@matf.bg.ac.rs}
%
\author{\firstname{Vladimir} \surname{Zekovi\'c}}
\email[]{vlada@matf.bg.ac.rs}
\affiliation{Department of Astronomy, Faculty of Mathematics, University of Belgrade, Studentski trg 16, 11000 Belgrade, Serbia}

\begin{abstract}
Prescriptions for electron injection into the diffusive shock acceleration process are required in many practical considerations of cosmic-ray astrophysics, particularly in modeling of the synchrotron emission of astrophysical sources. In particle-in-cell simulations of quasi-parallel magnetized collisionless shocks, we analyse the evolution of particle spectra. We find that in the later stages of shock evolution, the initially strong suprathermal part in the ion spectra fades, thus leaving the spectra composed of a Maxwellian and a power law. Once the electron and ion spectra flatten and become parallel, we find that the amounts of cosmic ray ions and electrons become similar.  We make the step towards relating the micro and macro-scale physics by applying this injection rule to Blasi's semi-analytical model of non-linear diffusive shock acceleration, in order to obtain the particle spectra and electron-to-proton ratio $K_{\mathrm{ep}}$ at high energies. By using shock jump conditions that include the electron heating, we find $K_{\mathrm{ep}}$ as a function of Mach number. For Mach number $\sim$ 100, our model finely reproduce the typically observed ratio for Galactic cosmic-rays $K_{\mathrm{ep}} \sim$ 1:100 in the test particle regime.\\


\end{abstract}

\maketitle

\section{\label{sec:intro}Introduction}

A promising mechanism for the acceleration of Galactic cosmic rays (CRs), mainly protons, based on the first order Fermi acceleration, the so-called diffuse shock acceleration (DSA), was developed independently by Axford et al (1977), Krymsky (1977), Bell (1978) and Blandford \& Ostriker (1978). In this test particle approach or linear DSA, it is assumed that the pressure of CRs is small, so that they do not modify the shock structure. If this is not the case, then we are talking about the non-linear DSA or CR back-reaction (see e.g. Drury 1983, Berezhko \& Ellison 1999, Malkov \& Drury 2001, Blasi 2002a,b).

Although the acceleration of cosmic-ray electrons is generally less understood than that of protons, there are great progresses in this field (see e.g. Malkov \& Drury 2001, Amano \& Hoshino 2007, Wieland et al. 2016, Bohdan et al. 2017, Diesing \& Caprioli 2019, Xu et al. 2020). A ``difficulty'' with electrons is that due to their much lower mass, they have much smaller gyroradii than protons. The size of the shock interface is however of the order of ion gyroradius (Caprioli \& Spitkovsky 2014, Kato 2015, Zekovi\'c 2019) which is clearly much larger than that of electrons. In order to reach the ion injection momentum (or energy) and leave the shock barrier, electrons need to be efficiently pre-accelerated or energized through the mechanisms that act inside the shock transition region.

On the other hand, assumptions concerning the injection of electrons into DSA processes are important in many practical aspects of cosmic-ray astrophysics, particularly in modeling of the synchrotron emission of astrophysical sources, such as supernova remnants (SNRs). In dealing with the radio evolution of SNRs, for the $\delta$-function-injection many authors use an assumption that the proton and electron momenta at injection are the same, with their number ratio $K_{\mathrm{ep}} \sim$ 1:100 (Berezhko \& Volk 2004, Pavlovi\'c 2017, Pavlovi\'c et al. 2018). While for protons, the non-relativistic value of $p_{\mathrm{inj}} = \xi p_{\mathrm{th}}$ with $\xi \sim 3-4 $ (Blasi et al. 2005) is easily achievable, the question ``Do electrons really need to reach such a high ion momentum to be injected?'' still remains. An answer to this question can be inferred from kinetic simulations.

In the case of quasi-perpendicular and exactly perpendicular shocks, electrons can at first experience the shock surfing acceleration (SSA). If shocks are weakly magnetized, it is shown by Amano \& Hoshino (2009) that electrons are reflected by large-amplitude electrostatic waves at the leading-edge of the shock transition region, rather than being reflected by the cross-shock electrostatic potential. Matsumoto et al. (2015) has shown that efficient electron energization can occur during turbulent magnetic reconnection. Upstream electrons can collide with reconnection jets and magnetic islands and thus experience the first-order Fermi acceleration. The contribution of the both mechanisms, SSA and magnetic reconnection, and their strong dependence on ion-to-electron mass ratio and Alfv\`en Mach number ($M_{\rm A}$), are considered by Bohdan et al. (2019). Therein, it is shown that SSA at the shock foot and the second order Fermi mechanism at the shock ramp could be dominant for lower mass ratios and $M_{\rm A}$, while for higher (more realistic) mass ratios and $M_{\rm A}$ the magnetic reconnection is the dominant mechanism (Bohdan et al. 2020). Moreover, the novel mechanism of electron acceleration by whistler waves via cyclotron resonance, which is proposed by Katou \& Amano (2019) as a stochastic shock drift acceleration (SSDA), is recently confirmed by observations at the Earth's bow shock (Amano et al. 2020). The SSDA is able to produce the power-law starting from the lowest electron energies. The properties that the electron power-law starts right from the Maxwellian, and that electrons are injected into DSA by scattering on the waves driven by themselves in the upstream, are also observed in PIC simulations of quasi-perpendicular (Xu et al. 2020) and quasi-parallel (Park et al. 2015) shocks, and we also confirm it here by our PIC simulations.

In the case of quasi-parallel shocks, Kato (2015) has found that electrons are heated by the upstream waves, which may lead to their steady injection into DSA in the later stages. In fact, Park et al. (2015) and Guo \& Giacalone (2015) have shown that electrons are continuously energized by the shock-drift acceleration (SDA), and then accelerated by the combination of both, SDA and DSA. Once electrons achieve the injection momentum of ions, they continue to behave similar to ions and accelerate only through the DSA mechanism.

Therefore, the common property which is found at both, quasi-parallel and quasi-perpendicular shocks, is that even though electrons ``formally'' enter DSA when they reach injection momentum (or energy, as discussed therein) of ions, their non-thermal spectrum $f (p) \propto p^{-4}$ goes all the way down to the electron thermal distribution. This was the motive to search for the injection momentum of electrons at quasi-parallel shocks, and to use it as a recipe in our model of non-linear DSA, that was conceptually considered in Arbutina \& Zekovi\'c (2019).

In Sec.~\ref{sec:PIC}, we present results of our PIC simulations and discuss the possible underlaying physical processes that in the later stages produce the observed, nearly equal amounts (number of CR particles relative to the total number of particles) of non-thermal ions and electrons $\eta_{i,e} = N_{i,e}^{\rm CR}/N_{i,e}$ ($i,e$ in the indices denotes ions and electrons, respectively). By using $\eta_e \sim \eta_p$ (the non-relativistic case; $p$ denotes protons) as a recipe in our model of non-linear DSA (which also includes constant electron heating ahead of the sub-shock) we derive the model equations in Sec.~\ref{sec:analysis}. We solve advection-diffusion equation numerically, and in the case of a real proton-to-electron mass ratio, magnetic field, and shock velocity, we obtain the particle spectra and the resulting $K_{\mathrm{ep}}$ dependence of the Mach number in Sec.~\ref{sec:results}. Finally, we compare the particle spectra obtained in our PIC runs with the spectra produced by our analytical model, and we discuss the physical implications that this model imposes.

\section{\label{sec:PIC}PIC Simulations of Quasi-parallel Shocks}

To find the relation between ion and electron injection, we ran kinetic simulations of initially parallel magnetized collisionless shocks with different ion--to--electron mass ratios and Mach numbers (as given in Table 1) by using the PIC code TRISTAN-MP (Spitkovsky 2005). We use both, fixed and expanding size simulation boxes of a rectangular shape in the $x-y$ plane, with periodic boundary conditions in the $y$ direction. The physical size of the fixed computational domain (in runs 1 and 3) is $25600 \times 25.6~c/\omega_{pe}$. In runs 2 and 4, we use the expanding simulation box that gradually enlarges ahead of the shock, as the moving plasma injector reaches the right wall of the domain. This allows us to make an optimal usage of the available computational resources, and to significantly extend the evolution of a shock with the mass ratio 16 (in comparison to run 1) by reaching the higher end times (compared to those in runs 1--3). At the same time, all particles and waves generated by the shock are preserved. In runs 1,2, and 4, the electron skin depth ($c/\omega_{pe}$) is resolved with 10 cells (with 5 cells in run 3), and each cell initially contains 8 particles (4 electrons and 4 ions). The noise is reduced by filtering particle contribution to the current 32 times per timestep. The end times of the simulation runs are given in Table 1. In order to prolong the simulation, we reduce the width of the simulation box to $\sim 2.5~c/\omega_{pe}$ (almost 1D) in run 4 (for $m_i/m_e = 16$). Sironi \& Spitkovsky (2011) have shown that a mass ratio $m_i/m_e = 16$ is large enough to separate the ion and electron scales, and to capture the correct acceleration physics at quasi-parallel shocks. The simulation in this run reaches the end time $t \sim 1930~ \omega_{ci}^{-1}$ which is long enough for the shock to enter the quasi-equilibrium state. At this stage, the suprathermal part in particle spectra completely fades, leaving the spectra composed of a Maxwellian and a power law. Simulation runs 1 and 3 correspond to extremely high sonic Mach numbers $M_{\mathrm{S}}$ (implying a cold upstream plasma) while in runs 2 and 4 sonic Mach number is more realistic.

\begin{table*}[t!]
\begin{minipage}{\textwidth}
  \centering
  \caption{Parameters for each simulation run. Magnetization is defined as the ratio of magnetic to kinetic energy density $\sigma = B_0^2/(4\pi \gamma _0 n_{i} m_{i} c^2)$; $u_0$ is the shock velocity in the lab frame; $M_{\mathrm{A}}$ is Alfv\'enic Mach number; $M_{\mathrm{S}} \approx \sqrt{\frac{5}{3} \frac{k (T_e+T_i)}{m_i}}$ is sonic Mach number, where $T_e \approx T_i$ are the upstream plasma temperatures; w is the width of simulation box in the units of an ion inertial length $\lambda_i$; $t$ is the end time of the simulation run in units of $\omega_{ci}^{-1}$ and $\omega_{pe}^{-1}$ in the last two columns, respectively.}
 \vspace{0.1cm}
 {\small
   \begin{tabular}{@{\extracolsep{1.0mm}} l c c c c c c r r @{}}
   \hline
   Run & $m_{i}/m_{e}$ & $\sigma$  & $u_0$ [$c$] & $M_{\mathrm{A}}$ & $M_{\mathrm{S}}$ & w[$\lambda_i$] & $t[\omega_{ci}^{-1}]$ & $t[\omega_{pe}^{-1}]$ \\
   \hline
  1 & 16 & 0.6$\times 10^{-3}$ & 0.33 & 13 & 1800 & 6.4 & 250 & 4.1$\times 10^4$\\
  2 & 50 & 1.0$\times 10^{-3}$ & 0.33 & 11 & 40 & 7.2 & 370 & 8.3$\times 10^4$\\
  3 & 100 & $0.6$$\times$10$^{-3}$ & 0.33 & 13 & 1800 & 2.6 & 130 & 4.1$\times 10^4$\\
  4 & 16 & $0.6$$\times$10$^{-3}$ & 0.4 & 16 & 35 & $<1$ & 1930 & 3.2$\times 10^5$\\
  \hline
\end{tabular}\\
\vspace{0.0cm}
 }
 \end{minipage}
\end{table*}

\begin{figure*}[h!]
\includegraphics[width=0.88\textwidth,keepaspectratio]{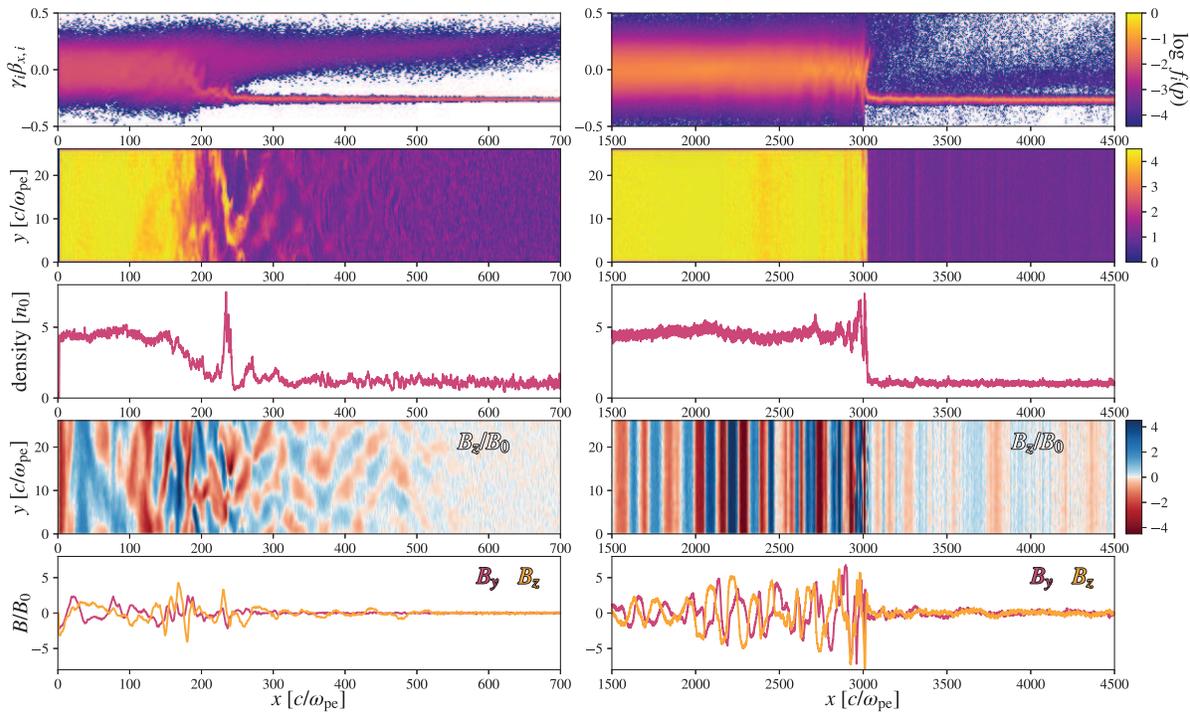}
\begin{center}
\caption{\label{fig:fields16} From top to bottom rows: the ion longitudinal phase space, density field, transversely averaged density profile, $B_z$ field, and transversely averaged perpendicular magnetic field profiles. The plots are given at early (left column) and late (right column) stages of the shock evolution in run 1 ($m_i/m_e=16$).}
\end{center}
\end{figure*}

\begin{figure*}[h!]
\includegraphics[width=0.88\textwidth,keepaspectratio]{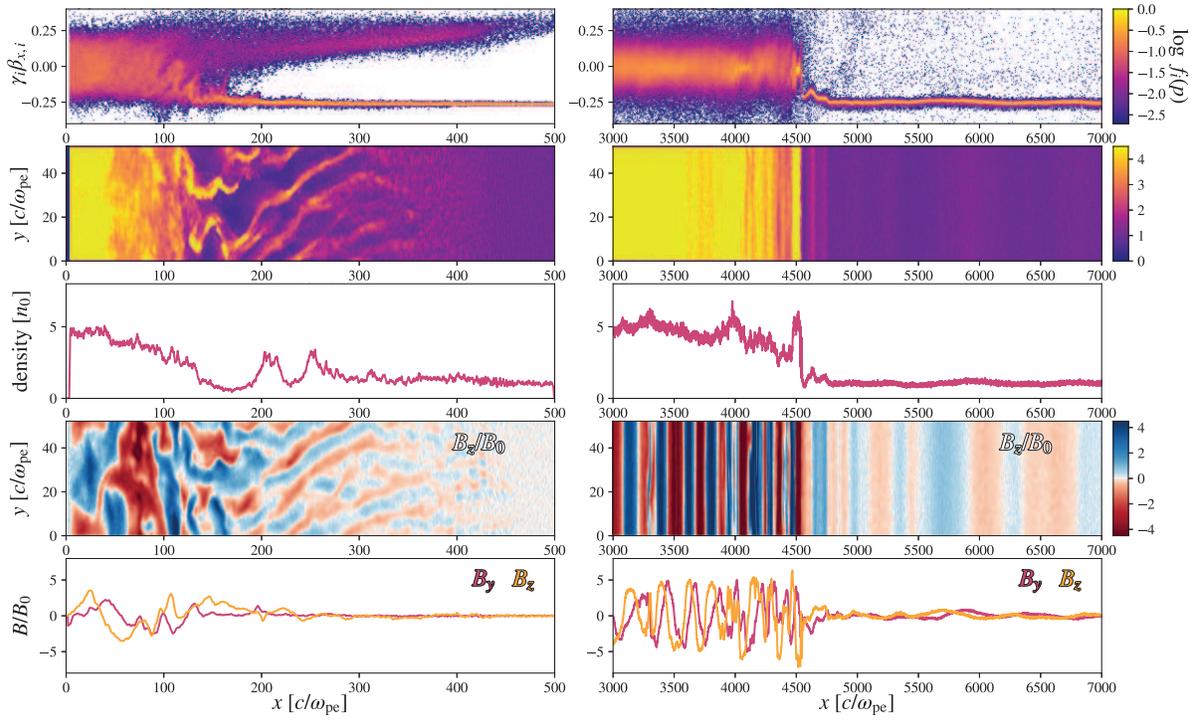}
\begin{center}
\caption{\label{fig:fields50} The same plots as in Fig.~\ref{fig:fields16}, but given for run 2 ($m_i/m_e=50$).}
\end{center}
\end{figure*}

\begin{figure*}[h!]
\includegraphics[width=0.88\textwidth,keepaspectratio]{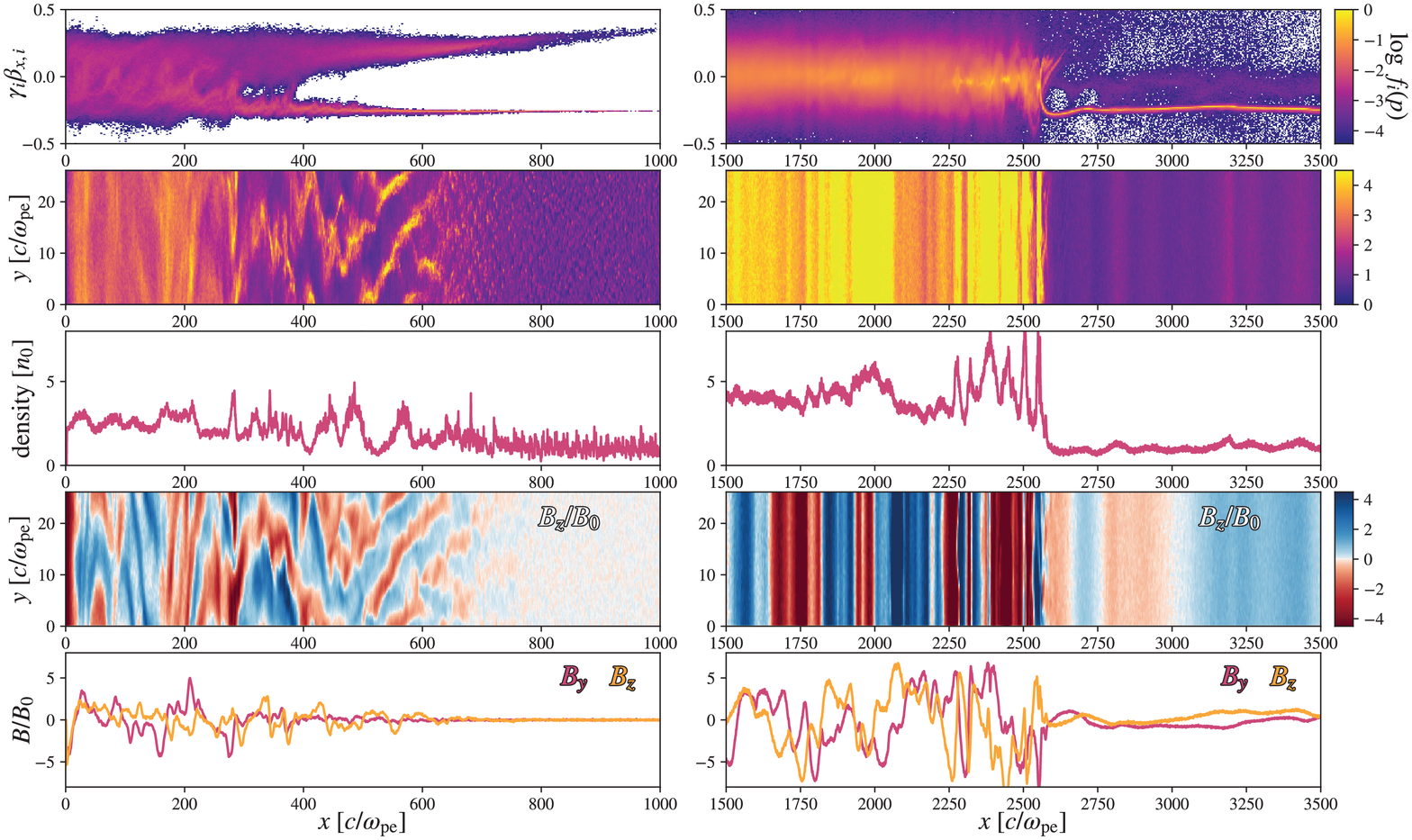}
\begin{center}
\caption{\label{fig:fields100} The same plots as in Figs.~\ref{fig:fields16}\&\ref{fig:fields50}, but given for run 3 ($m_i/m_e=100$).}
\end{center}
\end{figure*}

\begin{figure*}[h!]
\includegraphics[width=0.88\textwidth,keepaspectratio]{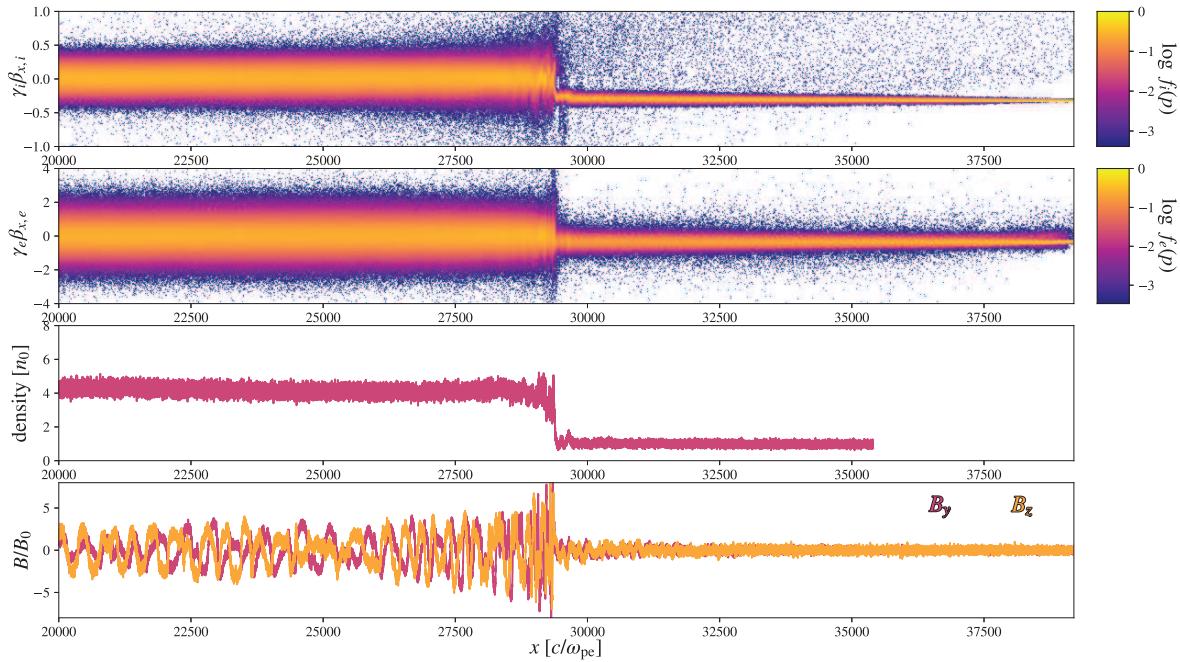}
\begin{center}
\caption{\label{fig:fields16xl} From top to bottom: ion and electron longitudinal phase spaces, transversely averaged density and perpendicular magnetic field profile. The plots are given at the end time of the long run 4 ($m_i/m_e=16$).}
\end{center}
\end{figure*}

\begin{figure*}[t!]
\includegraphics[width=0.32\textwidth,keepaspectratio]{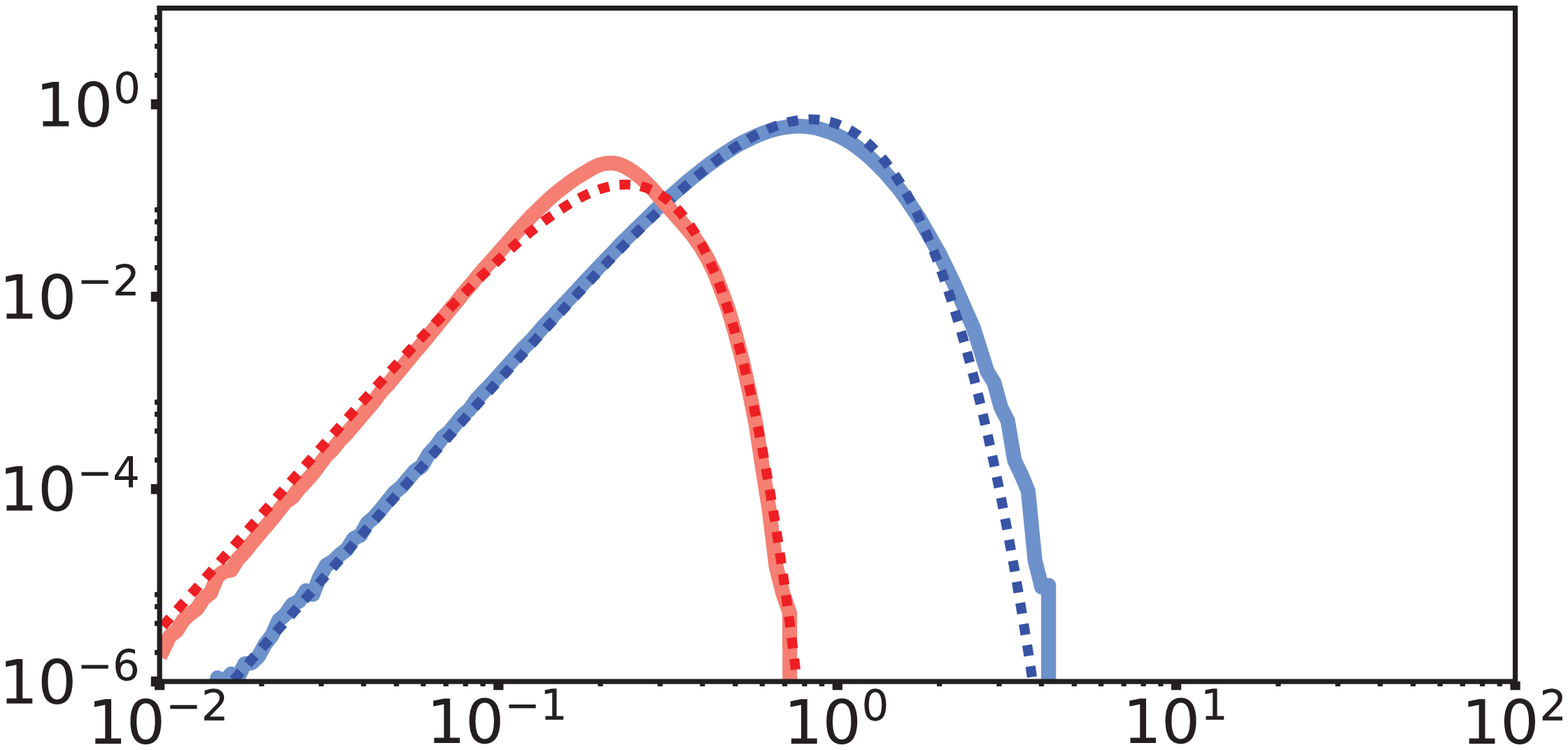}
\includegraphics[width=0.32\textwidth,keepaspectratio]{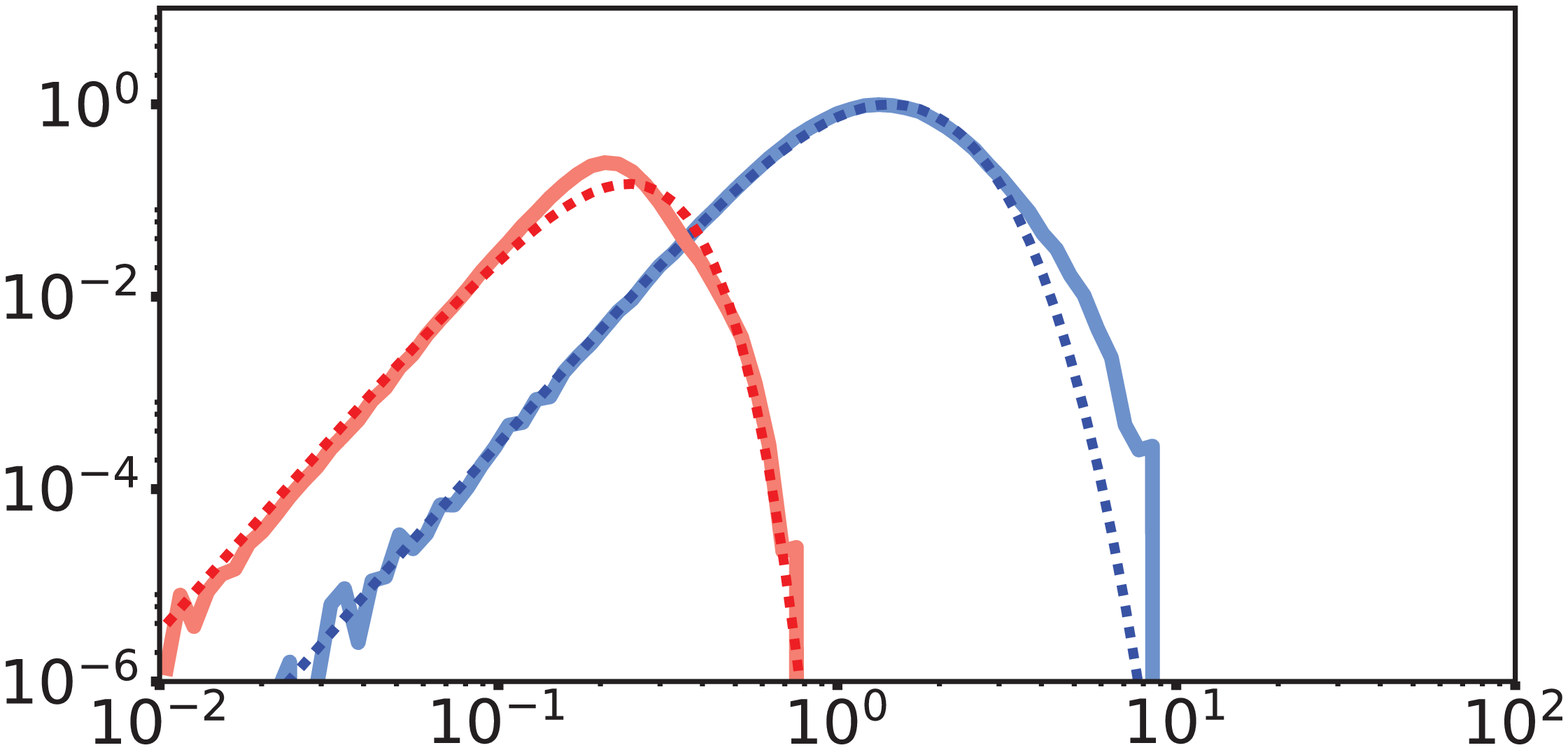}
\includegraphics[width=0.32\textwidth,keepaspectratio]{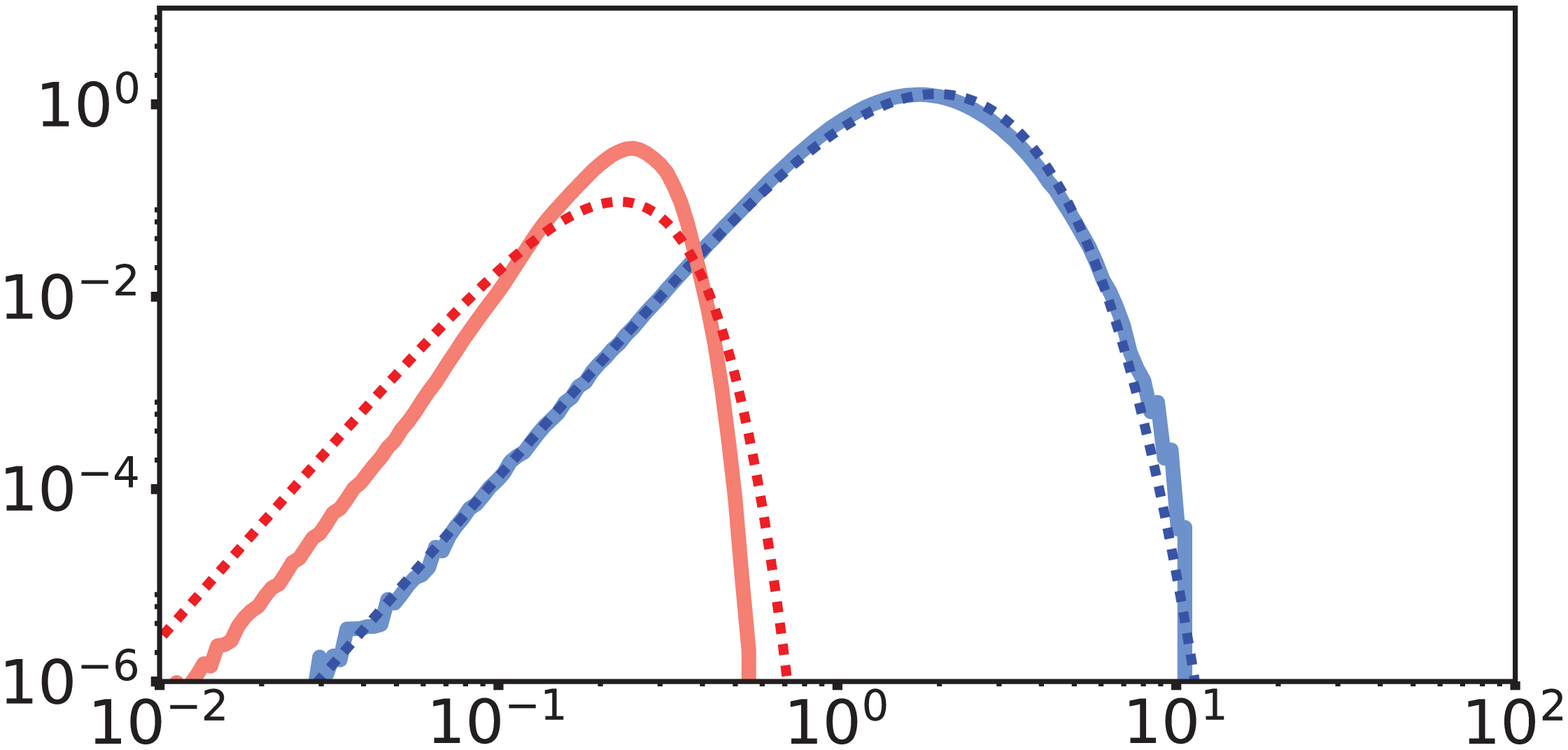}
\includegraphics[width=0.32\textwidth,keepaspectratio]{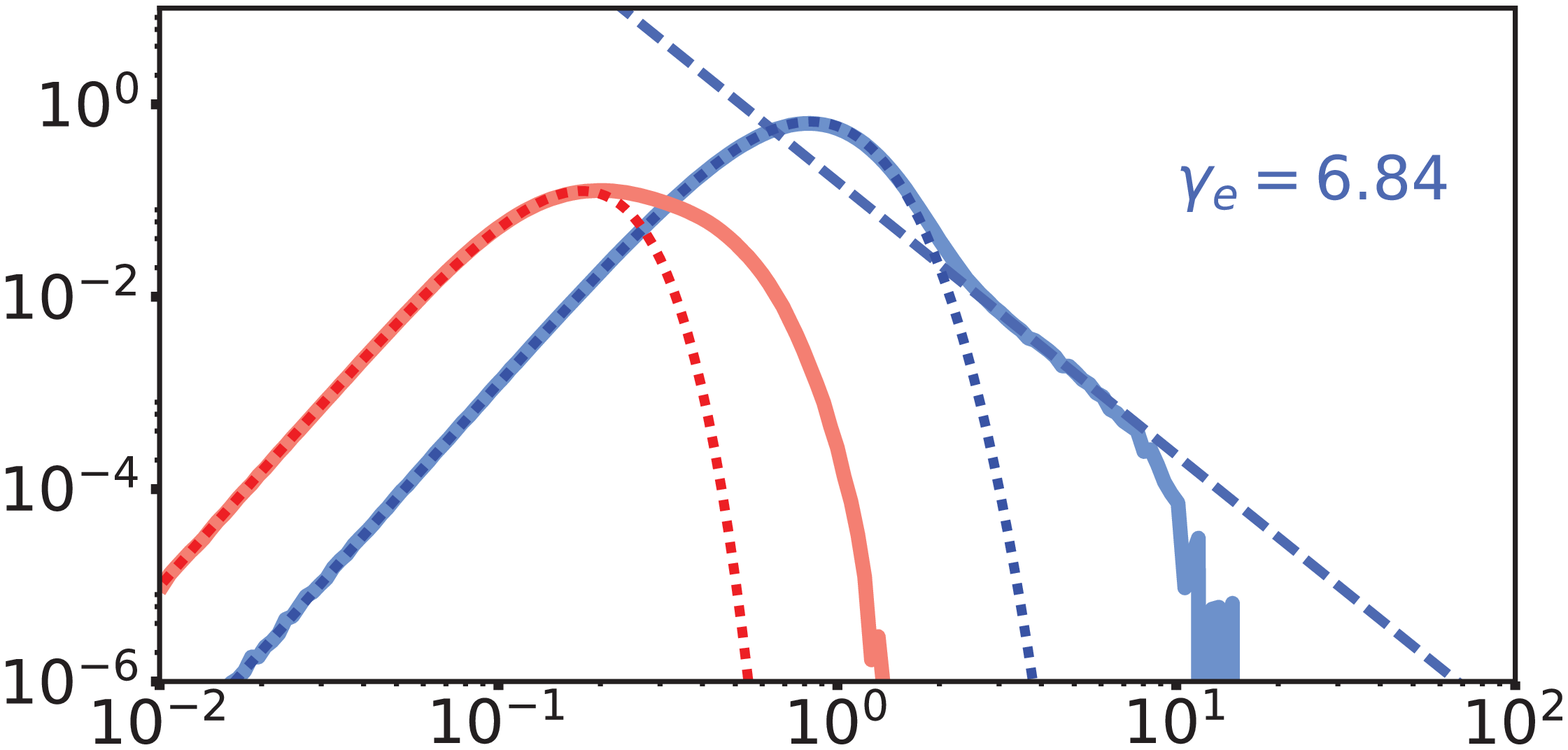}
\includegraphics[width=0.32\textwidth,keepaspectratio]{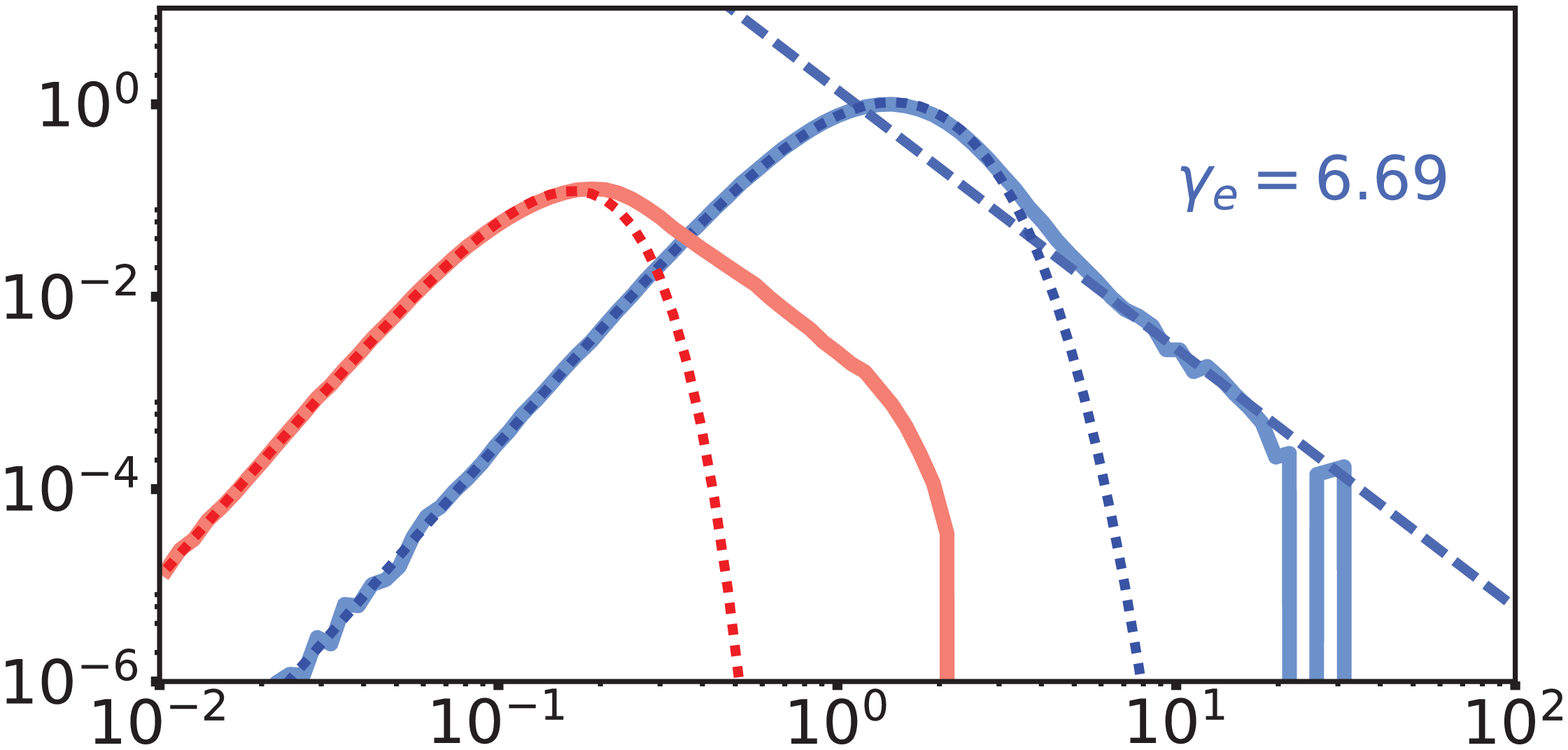}
\includegraphics[width=0.32\textwidth,keepaspectratio]{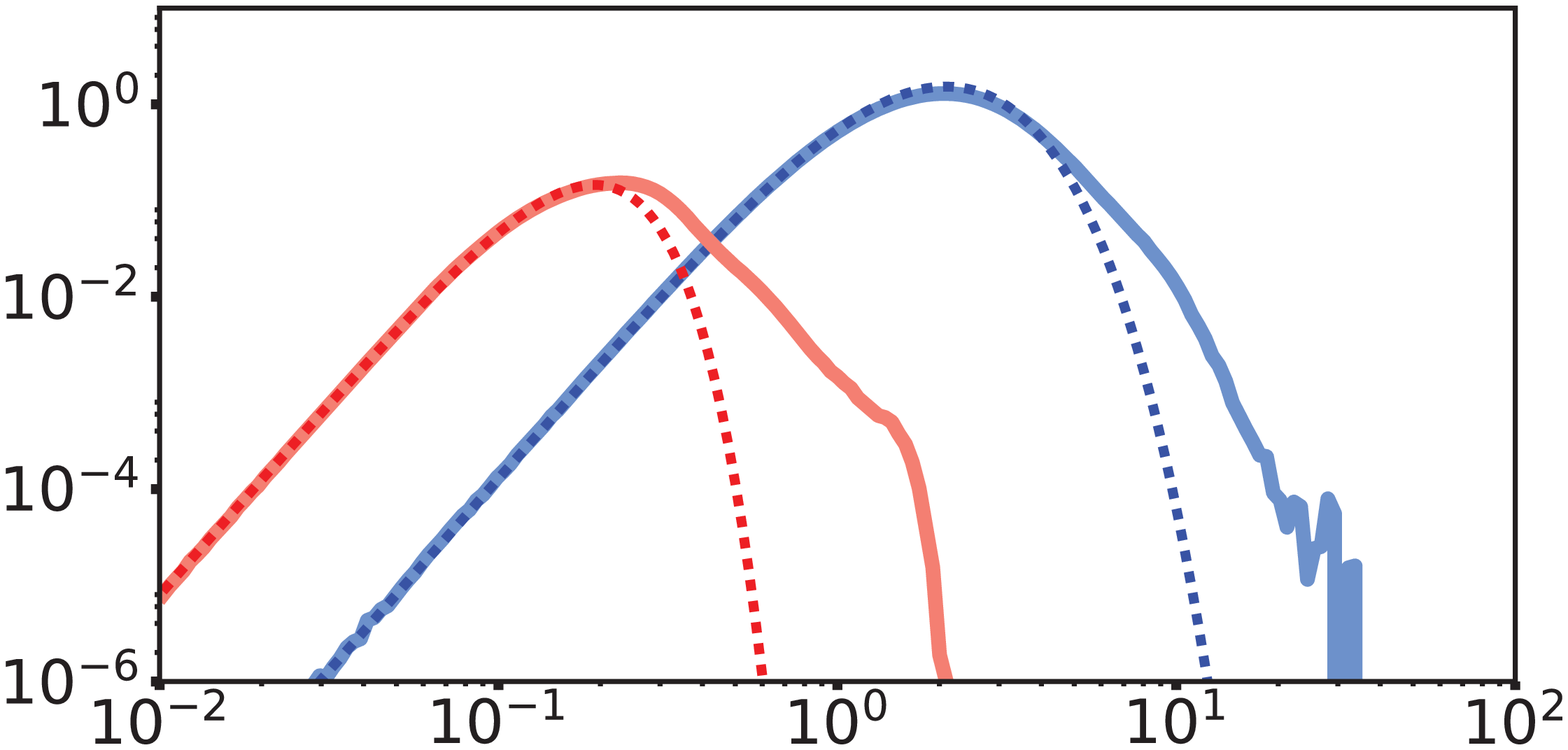}
\includegraphics[width=0.33\textwidth,keepaspectratio]{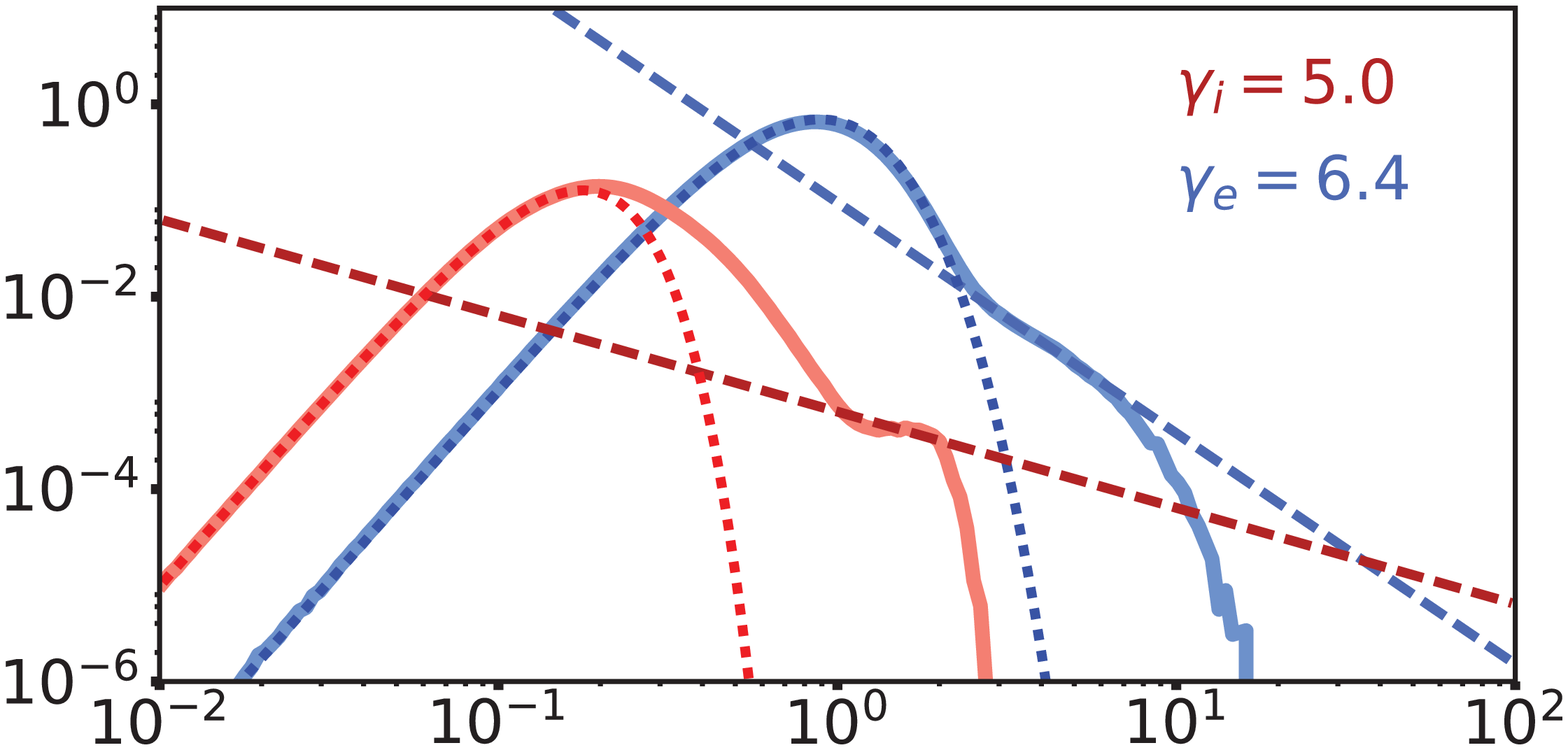}
\includegraphics[width=0.32\textwidth,keepaspectratio]{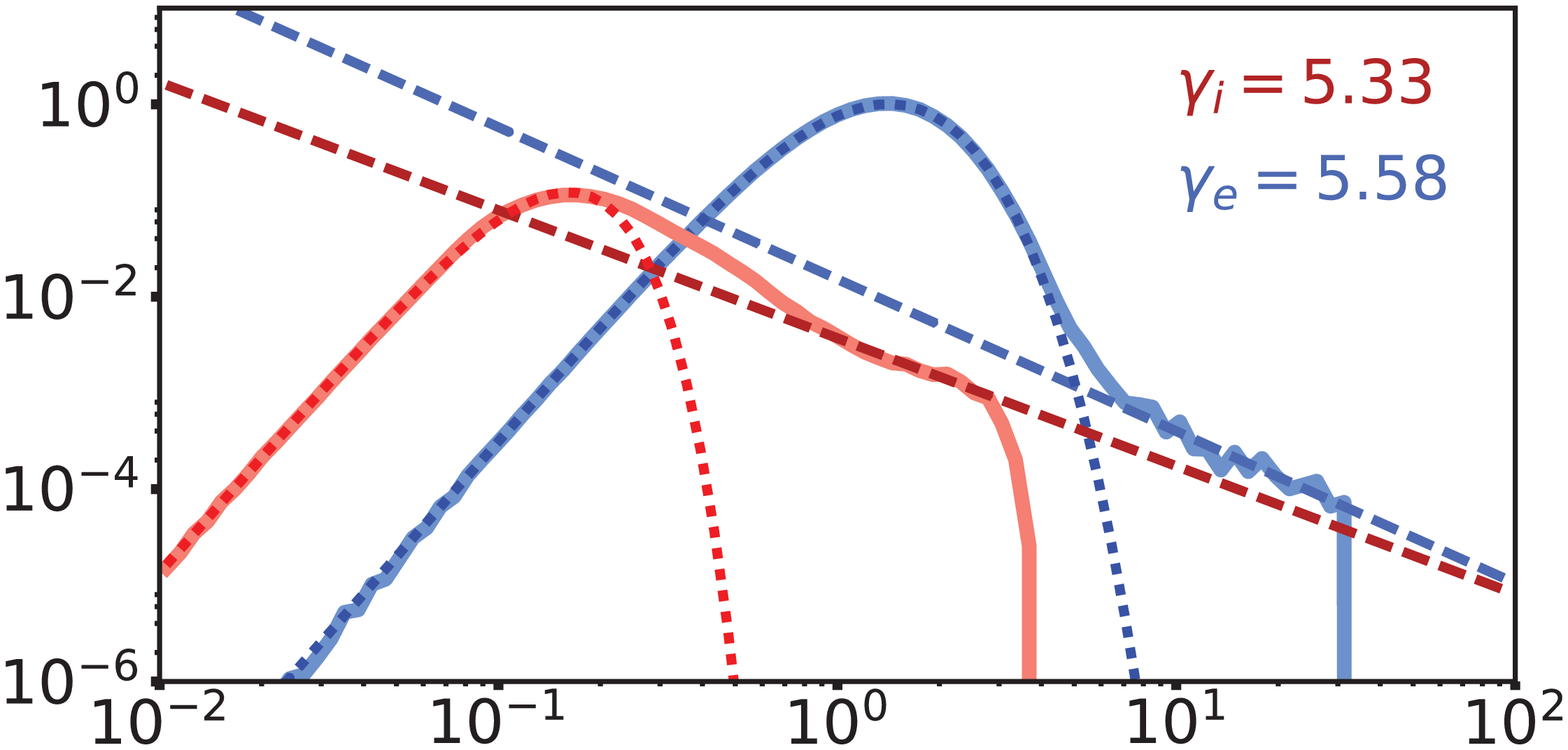}
\includegraphics[width=0.33\textwidth,keepaspectratio]{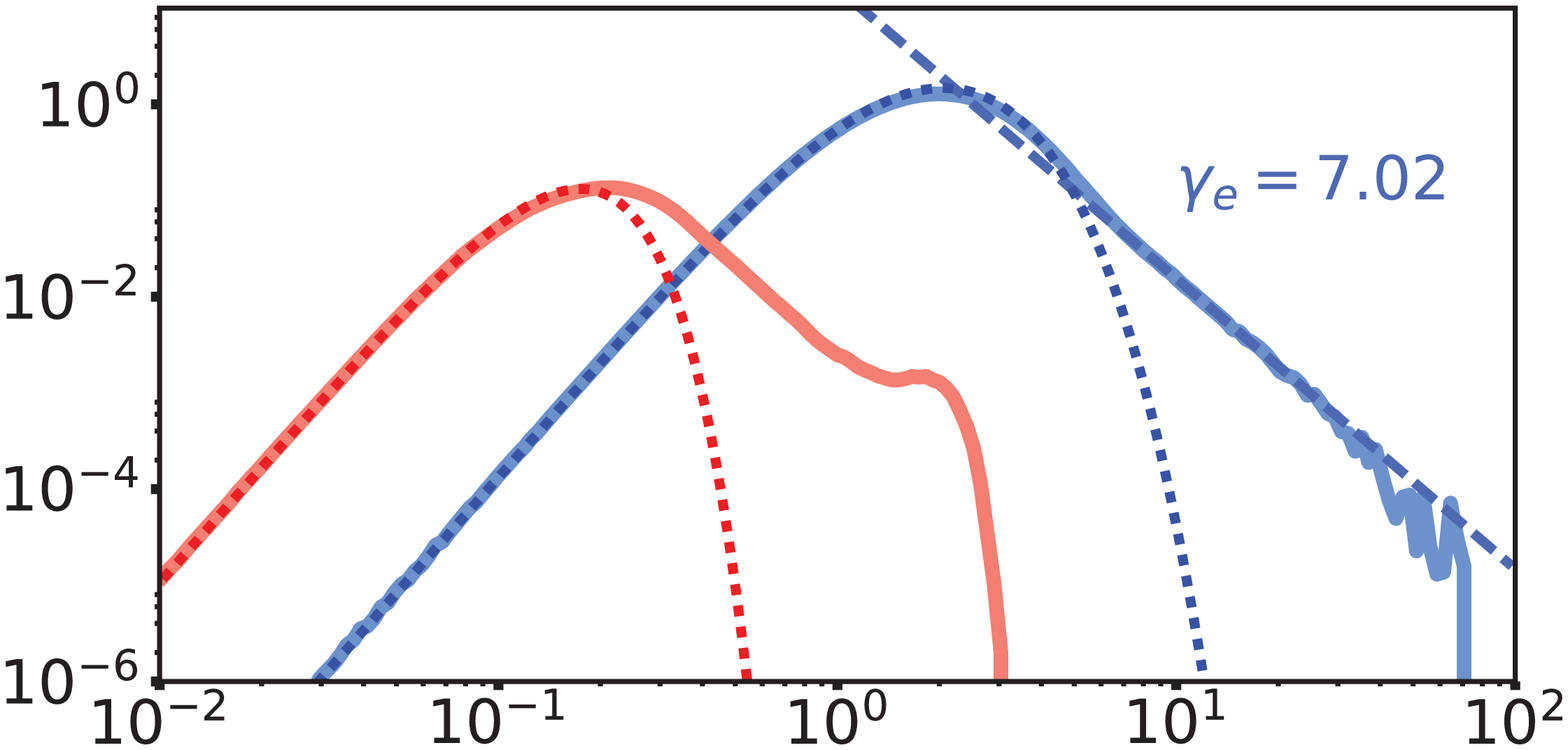}
\includegraphics[width=0.33\textwidth,keepaspectratio]{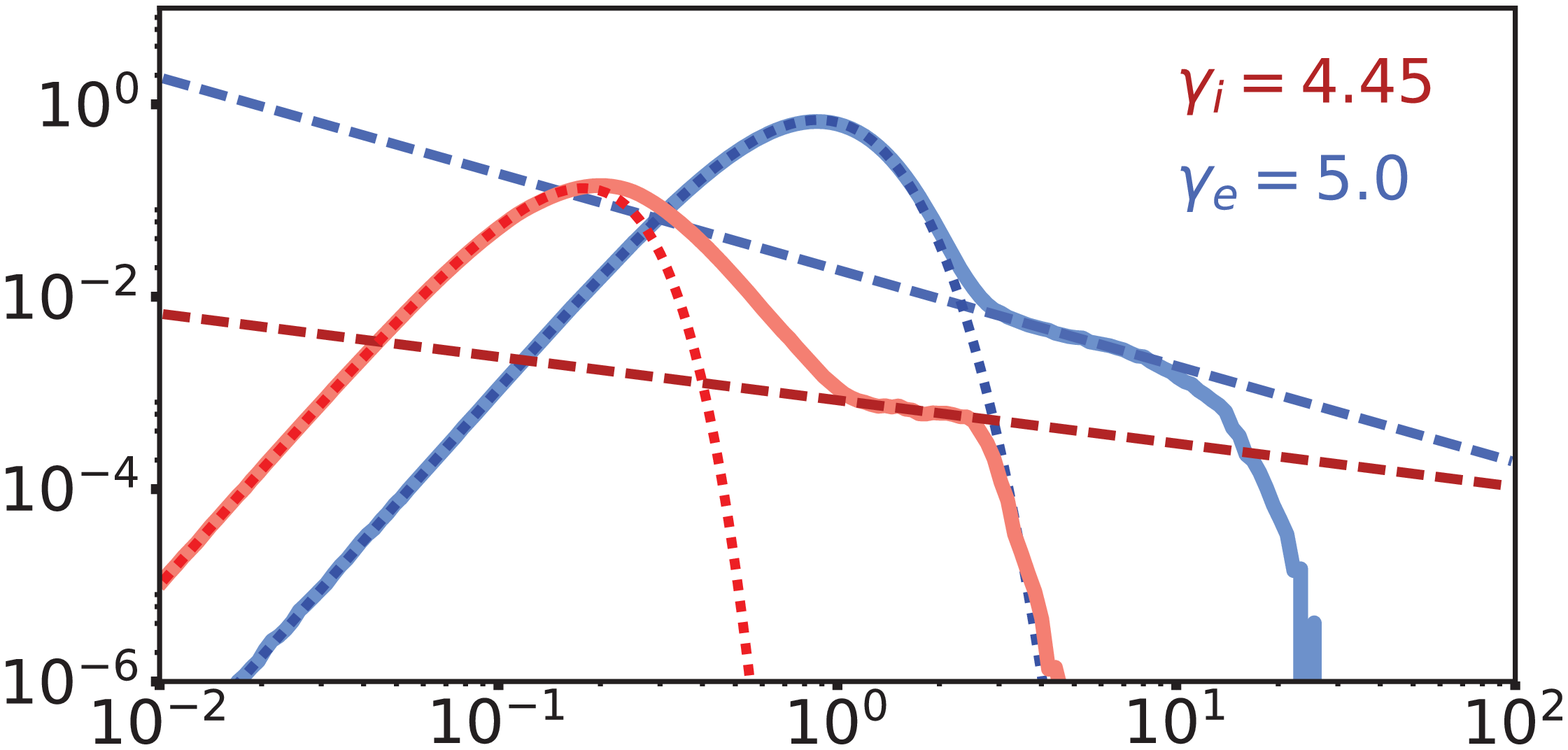}
\includegraphics[width=0.32\textwidth,keepaspectratio]{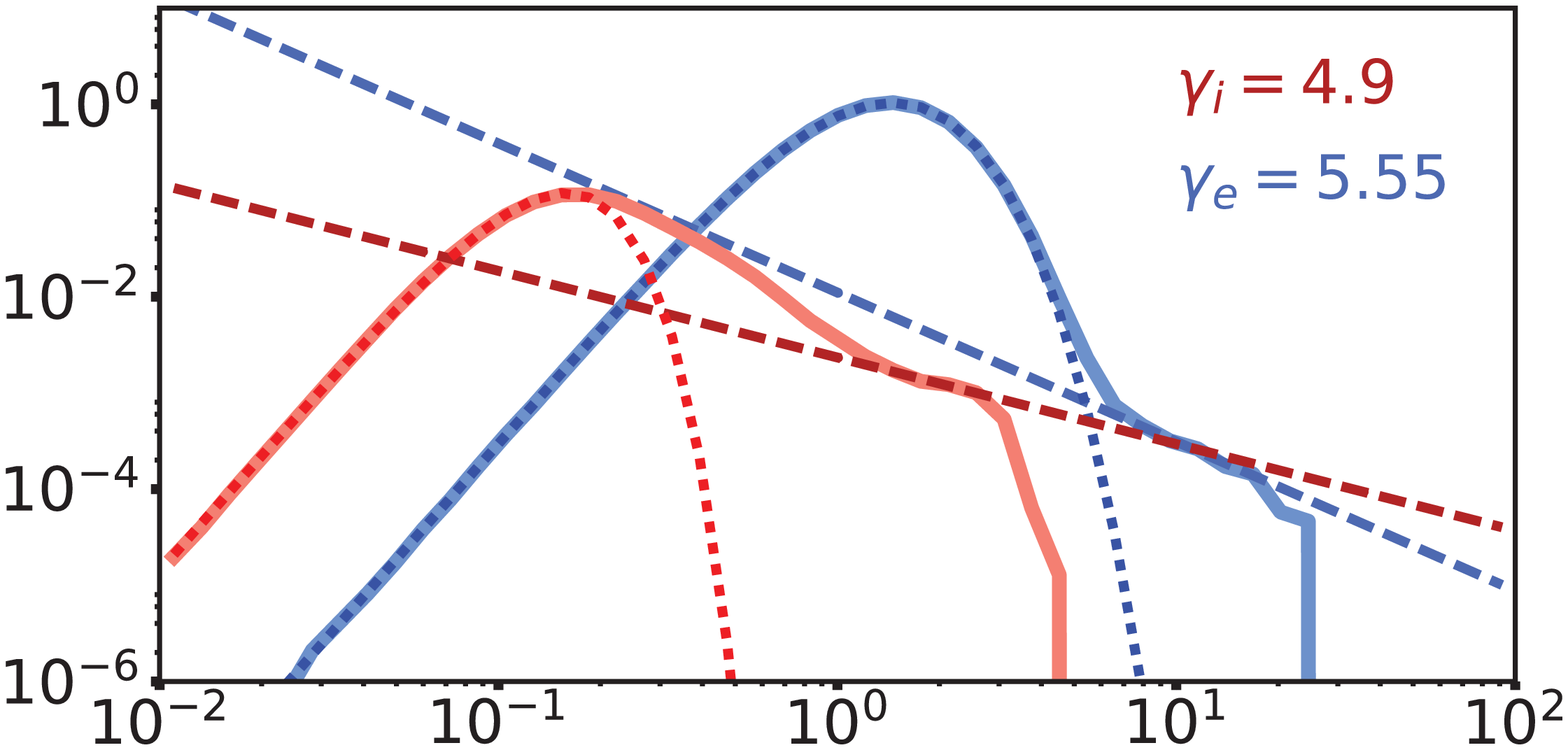}
\includegraphics[width=0.33\textwidth,keepaspectratio]{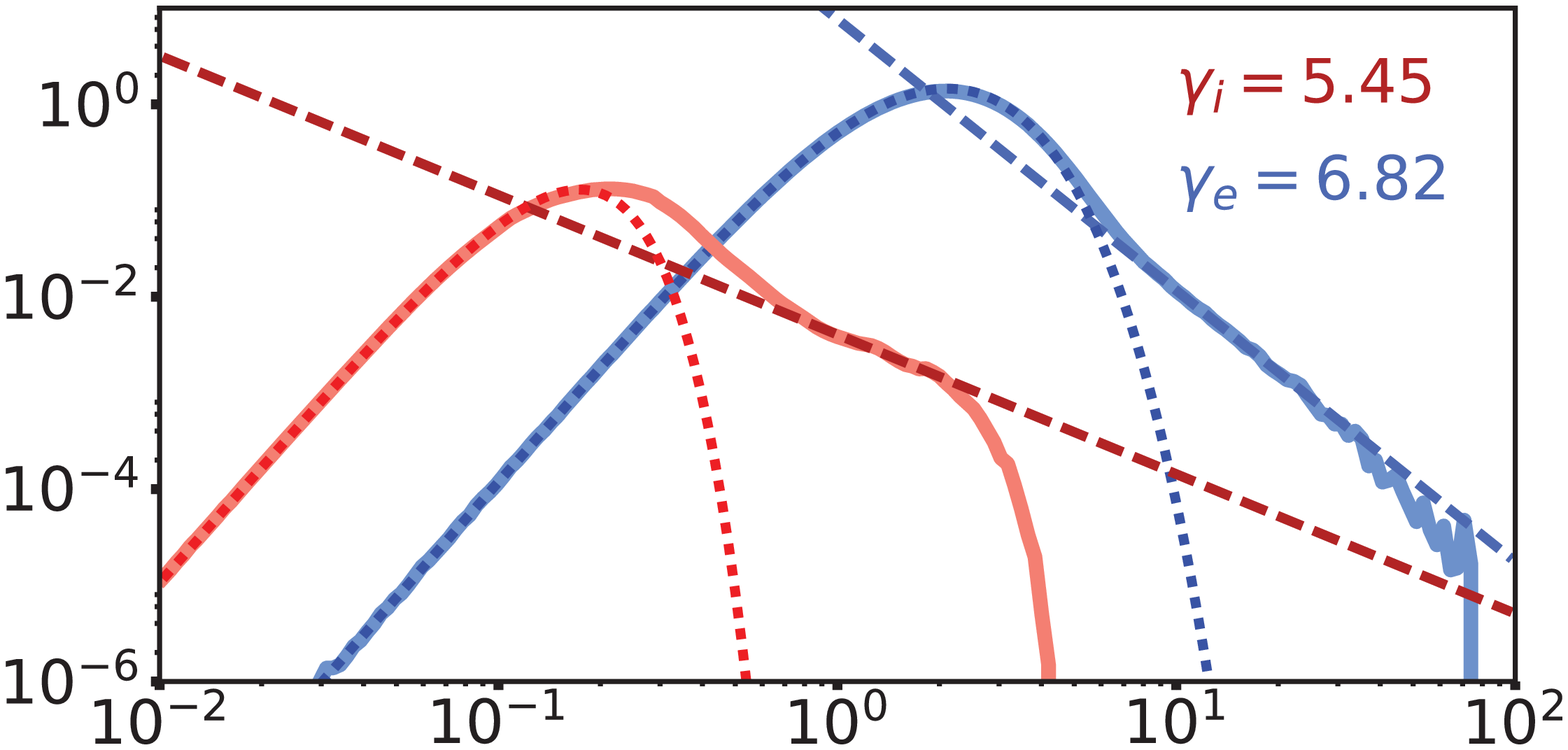}
\begin{center}
\caption{\label{fig:spectra} Ion (red) and electron (blue) downstream spectra at times $t \sim \{ 3, 16, 32, 38 \} \times 10^{3} \ \omega_{pe}^{-1}$, from top to bottom rows, respectively; in the simulation runs 1,2, and 3 (from left to right columns, respectively); with fitted Maxwelians and power-laws. On the horizontal axis, the momentum $p/mc$ is given. On the verical axis, the normalized $4 \pi p^4 f(p)$ is given, where $p$ is in the units of $mc$.}
\end{center}
\end{figure*}

In Figs.~\ref{fig:fields16}--\ref{fig:fields100} we show the ion phase space, density, and magnetic field plots for runs 1--3, captured while the shock was still forming, and, at the simulation end time. In Fig.~\ref{fig:fields16xl}, the particle phase spectra, density, and transverse magnetic field profiles are shown for run 4. At the very first stages in all our runs, the Weibel-type instability (Weibel 1959) grows faster (Crumley et al. 2019) than the resonant streaming instability (Zekovi\'c 2019). However, once the wave driven by the resonant instability grows to $\sim B_0$ in amplitude, it scatters the plasma flow and thus triggers the shock (re)formation which is further mediated by these modes. The upstream waves are seeded by the return current via non-resonant streaming instability (known as Bell's, or CR streaming instability; Bell 2004, Amato \& Blasi 2009). In the beginning of the shock evolution, beside the transverse Weibel modes, we also observe the local formation of quasi-stable magnetic regions ahead of the shock, that are composed of the two regions with opposite magnetic polarities. Such commonly observed 2D structures would correspond to the feet of a magnetic loop in 3D. They originate in the upstream, pass through the shock, and advect downstream, where they reconnect.

All these phenomena are related only to the very early stages, while in the later stages the shock is purely mediated by the Alfvenic-like modes (Bell's or CR streaming instability). Because these modes are shown to mediate the quasi-parallel shocks also in 1D runs (Kato 2015, Park et al. 2015), the relevant acceleration physics is still captured, despite the low transverse size (25 cells) of the simulation box in run 4. However, because in all our runs the box width is less than an ion gyro-radius ($r_{gi}$), the shock rippling as observed in e.g. Wieland et al. (2016) does not occur here. In reality, the shocks considered in this section may be highly influenced by the non-linear effects that appear on scales $\gtrsim r_{gi}$. Nevertheless, all the conclusions drawn in this section hold for the transverse scales up to $r_{gi}$, while we expect that for transverse scales larger than $r_{gi}$, the net effect on particle spectra should still remain the same in an average.

Although the initial number of particles per cell in all runs is relatively low, the noise level is below the level of modes that grow in the upstream (even in $\sim$1D run). Ion and electron distributions in the far upstream are given as Maxwellians. As the plasma enter the precursor, it becomes pre-heated by the upstream structure where it mixes with the non-thermal particles. At the shock transition, the plasma temperature increases, and the resulting spectra is composed of the Maxwellian, suprathermal part, and power-law.

In Fig.~\ref{fig:spectra} we show the downstream particle spectra obtained in our 2D simulation runs. We observe that ion supra-thermal bridge in each spectra fades over time, and non-thermal tail flattens at a rate that is similar in all runs. As this rate is measured in the units of $\omega_{ci}^{-1} = (\sqrt{\sigma^{-1} m_i / m_e})~\omega_{pe}^{-1}$, rather than $\omega_{pe}^{-1}$, the time scales at which shock evolves increases with the mass ratio. As a consequence, the more time is required for a shock to reach the evolutionary stages ($t \sim 1000\ \omega_{ci}$) at which changes in the transition between the thermal and the non-thermal component in spectra diminish. The spectra in individual rows, thus, correspond to different stages of the shock evolution given for each run.

\begin{figure}[h!]
\centerline{\includegraphics[width=0.64\textwidth,keepaspectratio]{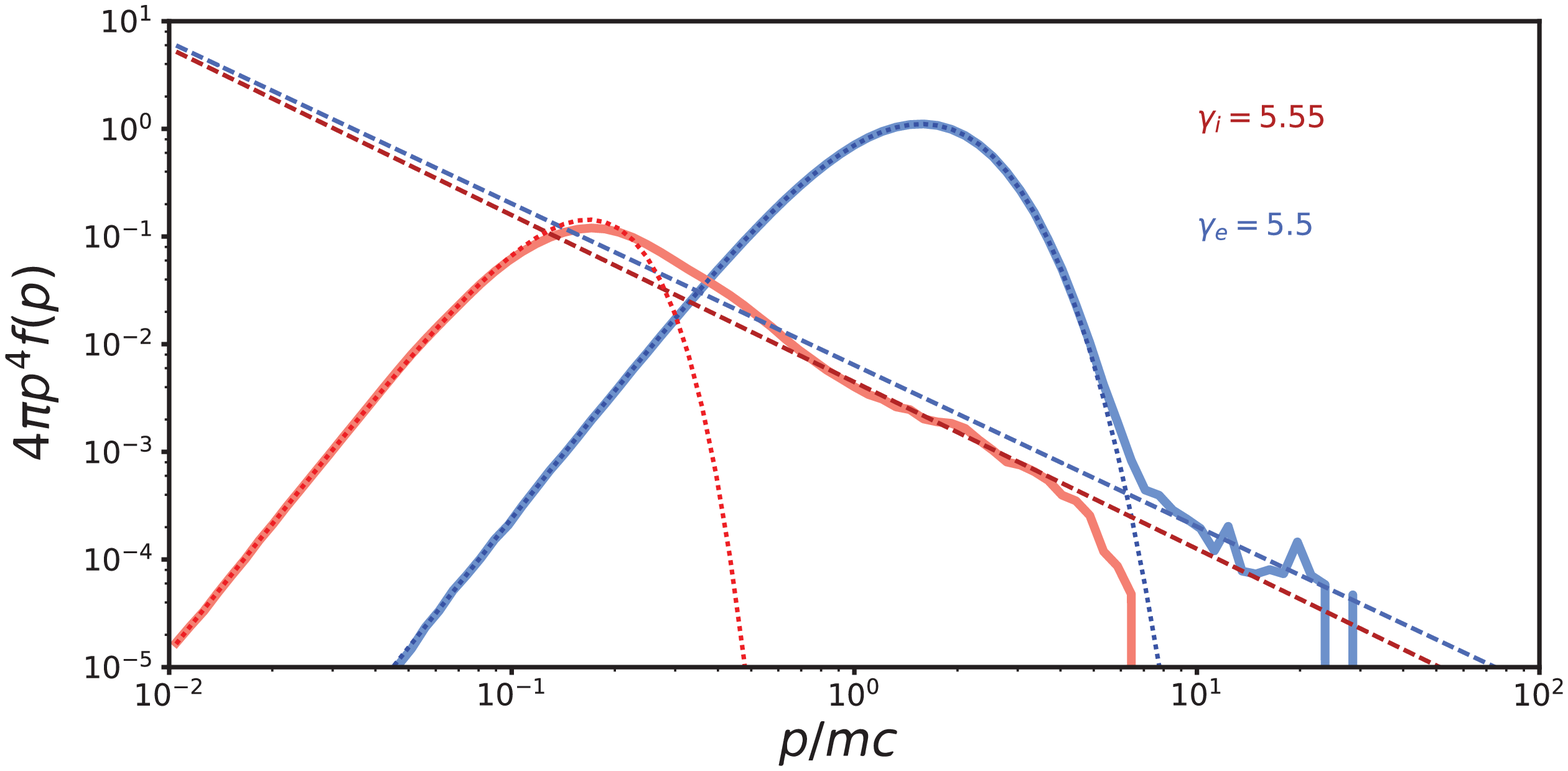}}
\begin{center}
\caption{\label{fig:end_spectra_m50} The downstream ion (red) and electron (blue) spectra at the end time of the simulation run 2 ($m_i / m_e = 50$). On the horizontal axis, the momentum $p/mc$ is given. On the verical axis, the normalized $4 \pi p^4 f(p)$ is given, where $p$ is in the units of $mc$.}
\end{center}
\end{figure}

The end spectra of the extended run 2 ($m_i/m_e=50$) is shown in Fig.~\ref{fig:end_spectra_m50}. Although the spectra in this run flattens (see Fig.~\ref{fig:spectra}), we get the sudden decrease in particle acceleration at $t \sim 6 \times 10^4 \omega_{pe}^{-1}$ followed by the gradual increase. This oscillation is related to the changes in the amplitude of the precursor wave. The similar oscillation is observed in Kato (2015) at comparable times. During this event, the electron acceleration almost entirely switches off, and ions start to build the intensive suprathermal bump in their spectrum. Once the particle acceleration is restored, we find that the non-thermal tails appear again and tend to become parallel.

\begin{figure}[h]
\includegraphics[width=0.97\textwidth,keepaspectratio]{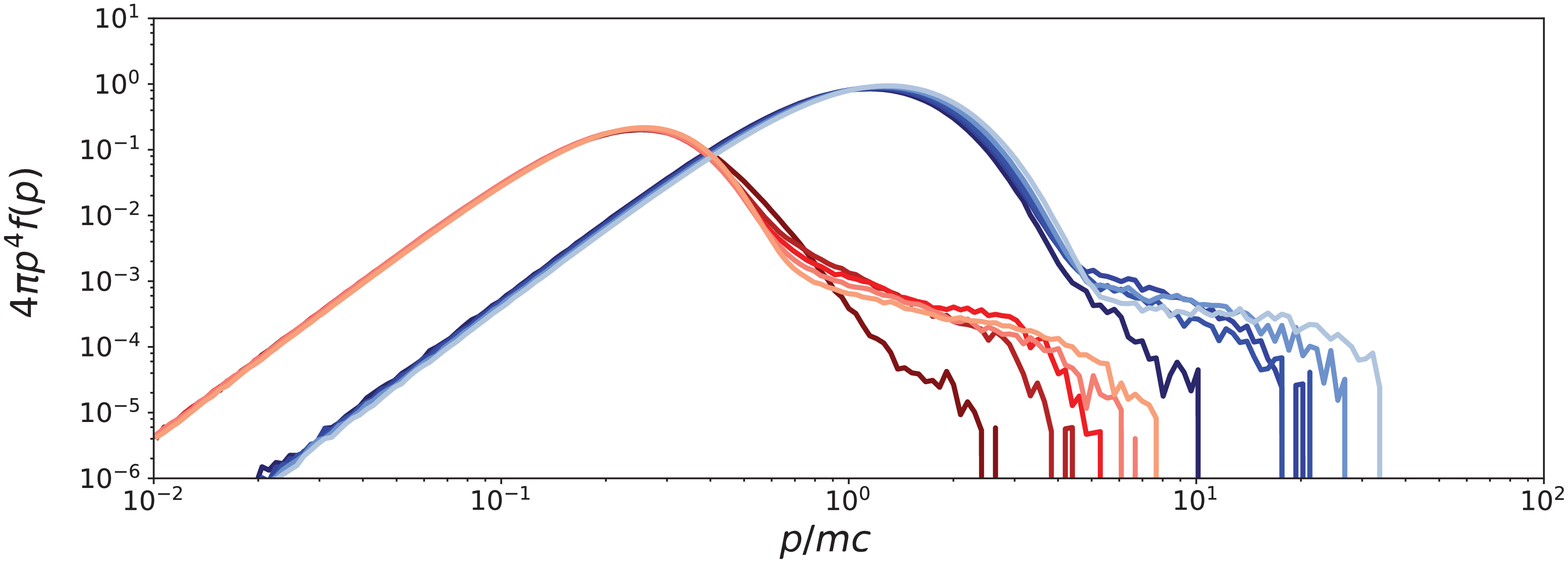}
\begin{center}
\caption{\label{fig:evolution_spectra} The evolution of the downstream ion (red coloring) and electron (blue coloring) spectra in the simulation run 4 ($m_i / m_e = 16$). Each color-coded line represents the spectra in a different time during the evolution. The darker lines correspond to earlier times, and the lighter lines correspond to later times. On the horizontal axis, the momentum $p/mc$ is given. On the verical axis, the normalized $4 \pi p^4 f(p)$ is given, where $p$ is in the units of $mc$.}
\end{center}
\end{figure}

In Fig.~\ref{fig:evolution_spectra}, we present the evolution of ion and electron spectra in the long run 4 ($m_i/m_e=16$). The diminishing of the supra-thermal part, followed by the slight increase in the downstream temperature, and flattening of the non-thermal part can be clearly seen. These properties are also shown in the evolution plot of ion spectrum in hybrid simulations of Caprioli \& Spitkovsky (2014). In our long run, we find that the advected suprathermal ions are losing their ``excess'' energy in the downstream and slightly heat the thermal plasma. During the shock evolution, this heating is observed as a small shift of the Maxwellian maximum towards higher energies. The very weak supra-thermal bump (which remains in the downstream spectra) is due to the still ongoing reformation process (as is apparent from Fig.~\ref{fig:fields16xl}). Simultaneously, the non-thermal tail flattens to a spectral index of $\sim 4$. Both these properties were observed in Caprioli \& Spitkovsky (2014), where at the later stages the non-thermal tail appears to be merged with the Maxwellian almost straightforwardly.

Conversely, in runs with the higher mass ratio, ion and electron spectra are clearly distinguished in the early stages of the shock evolution (Fig.~\ref{fig:spectra}). This difference in spectra is due to a much lower electron mass and, thus, much smaller gyroradius and faster dynamical evolution comparing to ions. Moreover, the wavelength of the upstream instability is related to the gyroradii of return ions (Zekovi\'c 2019). However, it is observed that electrons interact with the shock on their own scales through the various mechanisms (SSA, SSDA). It seems that electrons are not significantly influenced by the roughly constant escape probability which is imposed to ions during SDA cycles by the reforming shock barrier (Caprioli et al. 2014). It is observed in the case of quasi-perpendicular shocks that electrons are pre-accelerated by SSDA mechanism (Amano et al. 2020), which is expected to produce the power-law spectra starting from the lowest energies (the property observed also in PIC simulations by Xu et al. 2020). In the case of a quasi-parallel shock (Park et al. 2015) the electron non-thermal tail also emerges directly from the Maxwellian. Electrons are accelerated in combined, Fermi-like, SDA and DSA mechanisms. In the SDA cycles, electrons reflected by magnetic mirroring, gain energy in successive reflections between the two approaching structures -- the upstream and the shock fields (Xu et al. 2020). These reflections have the character of a Fermi I mechanism (acting almost like a ping-pong), and are localized to a narrow shock region. Due to that, we think that the Fermi-like SDA cycles (which produce the spiral-like trajectories in the electron phase space) at quasi-parallel shocks actually produce the power-law spectra which starts from the lowest electron energies (energy of electrons that are pre-heated in a precursor). This acceleration mechanism is almost equivalent to SSDA, but the microphysics behind the electron SDA is still to be shown. Because at quasi-parallel shocks, SDA seems to be the dominant pre-acceleration mechanism for electrons, the combination of SDA and DSA will thus not make the electron non-thermal spectra different from the power-law. As a consequence, we think that electrons have a very weak suprathermal bump, or may not have it at all. We observe that the electron spectrum in all our runs has a non-thermal tail which directly hits the Maxwellian. Such a sharp thermal--to--non-thermal transition appears in the late stages in the 1D (Park et al. 2015, Xu et al. 2020) as well as, in 2D (Crumley et al. 2019) runs.

The most important properties that we find in the late stages of our longest run are:

(i) the ion and electron non-thermal tails in the downstream spectra tend to become completely parallel over time, with the slope being $\sim 5$ in the near downstream, and its value $\to 4$ in the far downstream (similar to ion spectrum given in Caprioli \& Spitkovsky 2014);

(ii) the particles have their power-law emerging right from the Maxwellian at the point where injection--to--thermal momentum ratio ($\xi = p_{\rm inj}/p_{\rm th}$; $p_{\rm th} = \sqrt{2 m k T }$) implies the similar overall amounts of CR ions and electrons ($\eta_i \sim \eta_e$).

To show how these properties are related, we start with the CR differential distribution as given by

\begin{eqnarray}
\frac{dN^{\rm CR}(p)}{dp} = 4 \pi p^2 f(p) \approx \frac{\eta N (\Gamma - 1)}{p_{\rm inj}} \left( \frac{p}{p_{\rm inj}} \right)^{-\Gamma}, \nonumber
\end{eqnarray}

\noindent where $\Gamma$ is the energy index (the power-law slope), and $p_{\rm inj}$ is the momentum where the power-law tail hits the Maxwellian (we also call it the injection momentum), and $N$ is the total number of particles. If above distribution is integrated, we get that the total number of CRs is

\begin{eqnarray}
N^{\rm CR} = \int_{p_{\rm inj}}^{\infty} \frac{\eta N (\Gamma - 1)}{p_{\rm inj}} \left( \frac{p}{p_{\rm inj}} \right)^{-\Gamma} dp = \eta N, \ \ \left(\eta = \frac{N^{\rm CR}}{N} \right). \nonumber
\end{eqnarray}

In order to relate $\eta$ to $\xi$, the Maxwellian and power-law distributions are assumed to be equal at the point where $p = p_{\rm inj}$ and in the non-relativistic case we obtain

\begin{eqnarray}
\frac{4}{\sqrt{\pi}} N \frac{p_{{\rm inj}}^2}{p_{{\rm th}}^3} \cdot{\rm e}^{-\frac{p_{{\rm inj}}^2}{p_{{\rm th}}^2}} = \frac{\eta N (\Gamma - 1)}{p_{{\rm inj}}}, \nonumber \\
\frac{4}{\sqrt{\pi}} \frac{1}{\Gamma - 1} \ \xi^3 \cdot {\rm e}^{-\xi^2} = \eta. \nonumber
\end{eqnarray}

Therefore, the equal injection--to--thermal momenta ratios ($\xi_i \approx \xi_e$) and parallel power-laws ($\Gamma_i \approx \Gamma_e$) imply that amounts of injected ions and electrons are nearly equal ($\eta_{i} \approx \eta_e$). If an overall charge neutrality is assumed ($N_i = N_e = N$) then, the total number of CR ions is equal to the total number of CR electrons as well ($N_i^{\rm CR} = N_e^{\rm CR}$).

As specified by the previous dependence $\eta (\xi)$, this will only apply to non-relativistic thermal protons and electrons (whilst all of the simulated shocks are more or less relativistic). In a more general case of the relativistic Maxwellian gas, the distribution function is given by the Maxwell-J\"uttner distribution (Synge 1957)
\begin{equation}
f(p) = \frac{1}{4\pi m^3 c^3 \theta K_2 (1/\theta)} e ^{-\sqrt{1+ (p/mc)^2}/\theta}  ,
\label{eq:relmax1}
\end{equation}
where $\theta = \frac{kT}{m c^2}$ and $K_2(x)$ is the modified Bessel function of the second order. By requiring continuity of thermal and non-thermal distributions downstream one can obtain general relation between injection efficiency and parameter $\bar{\xi} = \frac{p_\mathrm{inj}}{m c}$
\begin{equation}
\eta = \frac{R-1}{3\theta K_2 (1/\theta)} \bar{\xi} ^3 e ^{-\sqrt{1+ \bar{\xi}^2}/\theta}  ,
\label{eq:relmax2}
\end{equation}
where $R = (\Gamma + 2) / (\Gamma -1)$ is the (sub)shock compression ratio. For large argument $K_2(x)$ can be approximated by (Abramowitz \& Stegun 1972)
\begin{equation}
K_2 (x) = \sqrt{\frac{\pi}{2x}} e^{-x} \Big( 1 + \frac{15}{8 x}+ \frac{105}{128 x^2} + \ldots \Big)  ,
\label{eq:relmax3}
\end{equation}
and keeping only the zeroth order term, one ends with Blasi et al. (2005) recipe
\begin{equation}
\eta = \frac{4(R-1)}{3\sqrt{\pi}} \xi^3 e^{-\xi^2} , \ \ \  \xi =\frac{p_\mathrm{inj}}{\sqrt{2 m k T}}.
\label{eq:relmax4}
\end{equation}
In the ultra-relativistic case, for small argument $K_2 (x) \approx 2/x^2$ (Abramowitz \& Stegun 1972), and
\begin{equation}
\eta = \frac{R-1}{6} \tilde{\xi}^3 e^{-\tilde{\xi}} , \ \ \  \tilde{\xi} =\frac{p_\mathrm{inj} c}{{k T}}.
\label{eq:relmax5}
\end{equation}

Electron-to-proton number (i.e. energy density ratios) has been discussed earlier, e.g. in Persic \& Rephaeli (2014), Merten et al. (2017), Park et al. (2015).
Generally, electron-to-proton ratio at high energies is
\begin{eqnarray}
K_{\mathrm{ep}} = \frac{\eta _e}{\eta _p} \Bigg(\frac{p_\mathrm{inj,e}}{p_\mathrm{inj,p}}\Bigg)^{\frac{3}{R-1}} &&= \frac{\eta _e}{\eta _p} \Bigg(\frac{m_\mathrm{e}}{m_\mathrm{p}} \frac{\bar{\xi}_\mathrm{e}}{\bar{\xi}_\mathrm{p}}\Bigg)^{\frac{3}{R-1}} \nonumber \\
&&=
\Bigg\{ \begin{array}{l}
   \frac{\eta _e}{\eta _p} \Big(\frac{{\xi}_\mathrm{e}}{{\xi}_\mathrm{p}}\Big)^{\frac{3}{R-1}} \Big(\frac{m_\mathrm{e}}{m_\mathrm{p}} \frac{T_\mathrm{e}}{T_\mathrm{p}}\Big)^{\frac{3}{2(R-1)}}, \hfill \theta _e, \theta _p \ll 1 , \\
    \frac{\eta _e}{\eta _p} \Big(\frac{\tilde{{\xi}}_\mathrm{e}}{\tilde{{\xi}}_\mathrm{p}}\Big)^{\frac{3}{R-1}} \Big(\frac{T_\mathrm{e}}{T_\mathrm{p}}\Big)^{\frac{3}{R-1}}, \hfill \theta _e, \theta _p \gg 1 .
 \end{array}
\end{eqnarray}

The caution needs to be taken here, because we refer to $p_{\rm inj,e/p}$ as the injection momentum. For protons it is very likely that $p_{\rm inj,p}$ is the momentum at which protons are indeed injected into DSA. However, the electron pre-acceleration mechanisms (SSDA for quasi-perpendicular, and Fermi-like SDA for quasi-parallel shocks) produce the power-law spectra, starting from the lowest electron energies. Due to that, it is more likely that the electron power-law starts inside the Maxwellian. Therefore, $p_{\rm inj,e}$ corresponds to the momentum at which electron power-law emerges from the Maxwellian, which does not necessarily means it is the power-law starting point.

In the following section we use the properties inferred from PIC simulations that (in the late stages) $\xi$ and $\Gamma$ imply similar injection fractions $\eta_{e,p}$, i.e. the total number of protons and electrons. Because in the run 4, electrons are relativistic, their $\xi$ will not match the ion one (although $\Gamma$ will be the same). However, in the non-relativistic case we have
\begin{eqnarray}
\xi_e \sim \xi_p, \nonumber \\
\Gamma_e \sim \Gamma_p, \nonumber \\
\eta_e \sim \eta_p. \nonumber
\end{eqnarray}

We apply Blasi's (2002a,b, 2004) semi-analytical model of non-linear DSA with the above parameters included, in order to obtain the particle spectra and electron-to-proton ratio at high energies $K_\mathrm{ep}$.

\section{\label{sec:analysis} Semi-Analytical Model of Shock Modification}

The details of the Blasi's semi-analytical model of non-linear DSA can be found in Blasi (2002a,b) (see also Blasi 2004, Blasi et al.~2005, Blasi et al.~2007, Amato \& Blasi 2005, Ferrand 2010, Pavlovi\'c 2018, and Uro\v sevi\'c et al. 2019).
We here give an overview of the derivation, which starts with the advection-diffusion equation and assumes that particles of a certain momentum $p$ will diffuse upstream [$0^-, -\infty$] to some distance
\begin{equation}
x_p = \frac{D(p)}{u_p},
\end{equation}
where $u_p$ is an average fluid velocity experienced by  particles with momentum $p$
\begin{equation}
u_p = u_1 - \frac{1}{f_0} \int_{-\infty}^{0^-} \frac{\mathrm{d} u}{\mathrm{d} x} f(x,p) \mathrm{d}x,
\label{eq:up}
\end{equation}
$D(p)$ is the diffusion coefficient assumed to be an increasing function of momentum, and $f(x,p)$ is the CR distribution function. Particles of momentum $p$ reach some $x_p$, and ``see" only part of a precursor in the velocity profile. Therefore, $u_p$ is interpreted as some typical fluid velocity at position $x_p$.
Blasi (2002a,b) shows that advection-diffusion equation can be transformed into:
\begin{equation}
\frac{1}{3} p \frac{\mathrm{d} f_0}{\mathrm{d} p} (u_2 - u_p) - f_0 \left(u_p+\frac{1}{3}
p \frac{\mathrm{d}u_p}{\mathrm{d}p} \right) + Q_0(p) = 0 ,
\label{eq:step1}
\end{equation}
where $f_0$ is the distribution function at the shock and $Q_0(p)$ is the so-called injection term. Eq. (\ref{eq:step1}) represents an ordinary linear differential equation that gives $f_0(p)$ is $u_p$ is regarded as known:
\begin{eqnarray}
f_0 (p) &=& \int_{p_0}^{p} \frac{\mathrm{d}{\bar p}}{{\bar p}}
\frac{3 Q_0({\bar p})}{u_{\bar p} - u_2} \exp\left[-\int_{\bar p}^p
\frac{\mathrm{d}p'}{p'} \frac{3}{u_{p'} - u_2}\left(u_{p'}+\frac{1}{3}p'
\frac{\mathrm{d}u_{p'}}{\mathrm{d} p'}\right)\right] \nonumber \\
&=& \frac{3 R_{\rm{sub}}}{R_{\rm{sub}}-1} \frac{\eta n_{\rm{1}}}{4\pi p_{\rm{inj}}^3}
\cdot
\exp\left[-\int_{p_{\rm{inj}}}^p
\frac{\mathrm{d}p'}{p'} \frac{3}{u_{p'} - u_2}\left(u_{p'}+\frac{1}{3}p'
\frac{\mathrm{d}u_{p'}}{\mathrm{d} p'}\right)\right].
\label{eq:inje}
\end{eqnarray}
In the above equation, monochromatic injection of particles with momentum $p_{\rm{inj}}$: $Q_0(p) = \frac{\eta n_{\rm{1}} u_1}{4\pi p_{\rm{inj}}^2} \delta(p-p_{\rm{inj}})$ is assumed, where $\eta$ is the injection efficiency (amount of injected particles); the gas number density immediately upstream ($x=0^-$) is $n_{\rm{1}}=n_0 R_{\rm{tot}}/R_{\rm{sub}} = n_0 R_{\rm{prec}}$, where $n_0$ is ambient density,  $R_{\rm{sub}}=u_1/u_2$ is the compression at the subshock, $R_{\rm{tot}}=u_0/u_2$ is the total shock compression and $R_{\rm{prec}}=u_0/u_1=R_{\rm{tot}}/R_{\rm{sub}}$ is the compression in the precursor.

As already mentioned, Blasi's model of injection (Blasi et al.~2005) assumes that
\begin{equation}
p_{\rm inj} = \xi p_{\rm th,2},
\label{eq:inj1}
\end{equation}
where thermal momentum $p_{\rm th,2}=\sqrt{2m_{\rm p}kT_2}$, $T_2$ is downstream temperature and $\xi$ is an injection parameter that can be brought into relation to injection efficiency $\eta$ by requiring continuity of thermal (Maxwell) and non-thermal distribution downstream at $p_{\rm inj}$, that is $f_{\rm th}(p_{\rm inj}) = f_0(p_{\rm inj})$. From this condition injection efficiency is found as:
\begin{equation}
\eta = \frac{4}{3\sqrt{\pi}}(R_{\rm sub}-1) \xi^3 e^{-\xi^2}.
\label{eq:eta}
\end{equation}
With dimensionless average fluid velocity defined as $U(p)= U_p = u_p/u_0$, Eq. (\ref{eq:inje}) takes the form:
\begin{equation}
f_0 (p) = \left(\frac{3 R_{\rm{sub}}}{R_{\rm{tot}} U(p) - 1}\right)
\frac{\eta n_{\rm{1}}}{4\pi p_{\rm{inj}}^3}
\cdot \exp \left[-\int_{p_{\rm{inj}}}^p
\frac{\mathrm{d}p'}{p'} \frac{3R_{\rm{tot}}U(p')}{R_{\rm{tot}} U(p') - 1}\right].
\label{eq:inje1}
\end{equation}
\noindent For $U_p\equiv1$ ($R_{\rm{tot}} = R_{\rm{sub}} = R$), the test-particle solution $f_0 \propto p^{-3R/(R-1)}$ is recovered. The non-linearity of the problem comes from $U_p = U_p (p)$, and $f_0(p)$ thus depends on the velocity profile $U_p$ through Eq. (\ref{eq:inje1}). However, $U_p$ itself depends on $f(p)$ in a non-linear fashion.

The $U_p$ is found (Blasi 2002a,b), by using the momentum conservation equation that relates quantities far upstream ($x\to-\infty$) with the quantities at $x_p$ (point reached by particles with momentum $p$), where fluid velocity is $u_p$:
\begin{equation}
 \rho_p u_p^2 +  P^\mathrm{p}_{\mathrm{th},p} + P^\mathrm{e}_{\mathrm{th},p} + P^\mathrm{p}_{\mathrm{CR},p} + P^\mathrm{e}_{\mathrm{CR},p} + P_{\mathrm{w},p} = \rho_0 u_0^2 + 2 P_{\mathrm{th},0} + P^\mathrm{p}_{\mathrm{CR},0} + P^\mathrm{e}_{\mathrm{CR},0} +
P_{\mathrm{w},0} ,
\label{eq:cont}
\end{equation}
as well as the mass conservation
\begin{equation}
\rho_0 u_0 = \rho_p u_p.
\label{eq:mass}
\end{equation}
In the above equations $\rho$ is the density, $P_{\rm{th}}$ the thermal pressure, $P_{\rm{CR}}$ the non-thermal CR pressure and $P_{\rm{w}}$ the pressure of plasma's hydromagnetic waves. The pressure equilibrium (and the thermal equilibrium) of the interstellar medium (ISM) protons and electrons, is assumed.

In the case of Alfven heating of plasma Berezhko \& Ellison (1999) suggested:
\begin{equation}
\frac{P^\mathrm{p}_{\mathrm{th},p}}{P_{\mathrm{th},0}} = U_p^{-\gamma}\left[1 + \zeta(\gamma -1)\frac{M_{\mathrm{S},0}^2}{M_{\mathrm{A},0}}(1-U_p^{\gamma})\right],
\label{eq:bif99}
\end{equation}
where $M_{\mathrm{S},0}=\frac{u_0}{c_{\mathrm{S},0}}$ is the Mach's number, $c_{\mathrm{S},0}=\sqrt{\gamma P_{\mathrm{th},0}/ \rho_0}$ ambient sound speed (for protons),  $M_{\mathrm{A},0}=u_0/\upsilon_{\mathrm{A},0}$ the Alfven-Mach number, with $\upsilon_{\mathrm{A},0}$ being the Alfven speed. The Alfven heating parameter $0\leq \zeta \leq 1$ was introduced later by Caprioli et al. (2009). For $\zeta =0$, an adiabatic approximation is obtained with no Alfven heating. For $\zeta =1$, there is an efficient heating, but without magnetic field amplification. For thermal electrons, adiabatic conditions are assumed

\begin{equation}
\frac{P^\mathrm{e}_{\mathrm{th},p}}{P_{\mathrm{th},0}} =
\left(\frac{\rho_p}{\rho_0}\right)^{\gamma} = \left(\frac{u_0}{u_p}\right)^{\gamma} = U_p^{-\gamma},
\label{eq:adiab}
\end{equation}

For the CR pressure in Eq. (\ref{eq:cont}), it is assumed that $P^\mathrm{p}_{\rm{CR},0}=P^\mathrm{e}_{\rm{CR},0}=0$. Since only the particles with momentum $\geq p$ can reach $x=x_p$, for protons is then found that:
\begin{equation}
P^\mathrm{p}_{\mathrm{CR},p}= \frac{4\pi}{3} \int_{p}^{p_{\mathrm{max, p}}} p^3 v(p) f^\mathrm{p}_0(p) \mathrm{d}p = \frac{4\pi}{3} \int_{p}^{p_{\mathrm{max, p}}} \frac{p^4 c^2}{\sqrt{m_\mathrm{p}^2 c^4 + p^2 c^2}} f^\mathrm{p}_0(p) \mathrm{d}p,
\label{eq:CR}
\end{equation}
where $v(p)$ is particle velocity and $p_{\mathrm{max}}$ is the maximum momentum reached by CR particles. This maximum momentum depends on relevant time-scales of acceleration, escape, and losses (Blasi et al. 2007). Similarly, for CR electrons
\begin{equation}
P^\mathrm{e}_{\mathrm{CR},p}=  \frac{4\pi}{3} \int_{p}^{p_{\mathrm{max, e}}} \frac{p^4 c^2}{\sqrt{m_\mathrm{e}^2 c^4 + p^2 c^2}} f^\mathrm{e}_0(p) \mathrm{d}p.
\label{eq:CRe}
\end{equation}
As with the CR pressures, the magnetic field or wave pressure is assumed to be $P_{\rm{w},0}=0$. In the precursor, for the (resonant) wave field Caprioli et al. (2009) suggested
\begin{equation}
\frac{P_{\mathrm{w},p}}{\rho_0 u_0^2} = \frac{1-\zeta}{4 M_{\mathrm{A},0}} U_p^{-3/2} (1-U_p^2),
\label{eq:alfaPw}
\end{equation}
where $U_p^{-3/2}$ is adiabatic compression of the field, and factor $1-\zeta$ account for the effects of Alfven heating in Eq. (\ref{eq:bif99}) -- the wave dumping (and thus the gas heating) must remain reasonably small for the magnetic field to be substantially amplified ($\zeta < 1$).

By setting $P_{\rm{CR},0}$ and $P_{\rm{w},0}$ to 0 in Eq. (\ref{eq:cont}), then dividing the equation by $\rho_0 u_{0}^2$, inserting Eqs. (\ref{eq:bif99}), (\ref{eq:CR}) and (\ref{eq:alfaPw}), and performing a derivative with respect to $p$, it is found that
\begin{eqnarray}
\frac{\mathrm{d}U_p}{\mathrm{d}p}\left[1 - \frac{U_p^{-(\gamma+1)}}{ M_{\mathrm{S},0}^2}\left(2 + \zeta(\gamma -1)\frac{M_{\mathrm{S},0}^2}{M_{\mathrm{A},0}}\right) - \frac{1-\zeta}{8 M_{\mathrm{A},0}} \frac{U_p^2+3}{U_p^{5/2}}\right] \nonumber \\
= \frac{4\pi c^2}{3\rho_0 u_0^2} \frac{p^4 f^\mathrm{p}_0(p)}{\sqrt{m_\mathrm{p}^2 c^4 + p^2 c^2}} + \frac{4\pi c^2}{3\rho_0 u_0^2} \frac{p^4 f^\mathrm{e}_0(p)}{\sqrt{m_\mathrm{e}^2 c^4 + p^2 c^2}}.
\label{eq:diff_cont}
\end{eqnarray}

For fixed Mach and Alfven-Mach numbers (that is velocity $u_0$ and parameters of the surroundings $\rho_0, P_0, B_0, \gamma$), $\eta$, $\zeta$,  $p_\mathrm{inj}$, $p_{\rm max}$, another relation must be found between $R_{\rm sub}, R_{\rm tot}, R_{\rm prec}$ (knowing that $R_{\rm tot} = R_{\rm sub}\cdot R_{\rm prec}$) from jump conditions at the subshock.
The CR pressure must be continuous across the subshock $P_{\mathrm{CR},1}=P_{\mathrm{CR},2}$, while for the thermal pressures Vainio \& Schlickeiser (1999) derived a modified Rankine-Hugoniout jump conditions in the presence of plasma's hydromagnetic waves
\begin{equation}
\frac{P^\mathrm{p}_{\rm th,2}}{P^\mathrm{p}_{\rm th,1}}=\frac{(\gamma+1)R_{\rm sub}-(\gamma-1)\left[1-(R_{\rm sub}-1)\Delta\right]}{(\gamma+1) - (\gamma-1)R_{\rm sub}},
\label{eq:vainio}
\end{equation}
where
\begin{equation}
\Delta = \frac{R_{\rm sub}+1}{R_{\rm sub}-1}\frac{[P_{\rm w}]^2_1}{P_{\rm th,1}}-\frac{2R_{\rm sub}}{R_{\rm sub}-1}\frac{[F_{\rm w}]^2_1}{P_{\rm th,1} u_1},
\label{eq:delta}
\end{equation}
and $[P_{\rm w}]^2_1$, $[F_{\rm w}]^2_1$~~are jumps in magnetic field pressure and magnetic energy flux, respectively (we will use notation $[Y]^2_1 = Y_2-Y_1$).

Caprioli et al.~(2008, 2009) calculated $[P_{\rm w}]^2_1$ and $[F_{\rm w}]^2_1$ for the waves, by considering their transmission and reflection: $[P_{\rm w}]^2_1=(R_{\rm sub}^2-1)P_{\rm w,1}$, $[F_{\rm w}]^2_1=2(R_{\rm sub}-1)P_{\rm w,1} u_1$, which when inserted in Eq. (\ref{eq:vainio}) give:
\begin{equation}
\frac{P^\mathrm{p}_{\rm th,2}}{P^\mathrm{p}_{\rm th,1}}=\frac{(\gamma+1)R_{\rm sub}-(\gamma-1)\left[1-(R_{\rm sub}-1)^3\frac{P_{\rm w,1}}{P^\mathrm{p}_{\rm th,1}}\right]}{(\gamma+1) - (\gamma-1)R_{\rm sub}}.
\label{eq:vainio2}
\end{equation}

Momentum conservation equation for protons at the subshock:
\begin{equation}
\rho_1 u_1^2 + P^\mathrm{p}_{\mathrm{th,1}} + P^\mathrm{p}_{\mathrm{CR,1}} + P_{\mathrm{w},1}   =  \rho_2 u_2^2 + P^\mathrm{p}_{\mathrm{th,2}} + P^\mathrm{p}_{\mathrm{CR,2}} +
P_{\mathrm{w},2},
\label{eq:subshock_cont}
\end{equation}
can then be transformed to
\begin{equation}
\frac{\rho_1 u_1^2}{P_{\rm w,1}}\frac{R_{\rm sub}-1}{R_{\rm sub}} + \frac{P^\mathrm{p}_{\rm th,1}}{P_{\rm w,1}}  \left( \frac{P^\mathrm{p}_{\rm th,2}}{P^\mathrm{p}_{\rm th,1}}-1\right) + R_{\rm sub}^2-1=0.
\label{eq:subshock_cont1}
\end{equation}
The Mach's number ahead of the subshock is introduced as $M_{\rm S,1}=u_1/c_{\rm S,1}$, where  $c_{\rm S,1}=\sqrt{\gamma P^\mathrm{p}_{\rm th,1}/\rho_1}$ which can be related to $M_{\rm S,0}$ by using Eq. (\ref{eq:bif99}):
\begin{equation}
\frac{M_{\rm S,1}^2}{M_{\rm S,0}^2}=\frac{\rho_1 u_1^2}{\rho_0 u_0^2} \frac{P_{\rm th,0}}{P^\mathrm{p}_{\rm th,1}}=R_{\rm prec}^{-\gamma-1} \left[1 + \zeta(\gamma -1)\frac{M_{\mathrm{S},0}^2}{M_{\mathrm{A},0}}(1-R_{\rm prec}^{-\gamma})\right]^{-1}.
\label{eq:MS1}
\end{equation}
From Eqs. (\ref{eq:vainio2}) and (\ref{eq:subshock_cont1}), it is found that
\begin{equation}
M_{\rm S,1}^2 = \frac{2R_{\rm sub}}{(\gamma+1)-(\gamma-1)R_{\rm sub} - 2R_{\rm sub}P_{\rm w,1}^*\left[\gamma - (\gamma-2)R_{\rm sub}\right]},
\label{eq:MS12}
\end{equation}
where
\begin{equation}
P_{\rm w,1}^* = \frac{P_{\rm w,1}}{\rho_1 u_1^2} = R_{\rm prec} \frac{P_{\rm w,1}}{\rho_0 u_0^2} = \frac{1-\zeta}{4 M_{\rm A,0}}R_{\rm prec}^{5/2} (1-R_{\rm prec}^{-2}).
\label{eq:Pw1z}
\end{equation}

For a fixed $R_{\rm prec}$, Eq. (\ref{eq:MS12}) is quadratic in $R_{\rm sub}$:
\begin{equation}
2(\gamma-2)M_{\rm S,1}^2 P_{\rm w,1}^* R_{\rm sub}^2 - \left[2 + (\gamma-1+2\gamma P_{\rm w,1}^*)M_{\rm S,1}^2\right]R_{\rm sub} + M_{\rm S,1}^2(\gamma+1) = 0.
\label{eq:kvad}
\end{equation}
Positive root of this equation gives $R_{\rm sub}$ as a function of $M_{\rm S,1}$ and $P_{\rm w,1}^*$, and consequently $R_{\rm tot}$. Therefore, the compression in the precursor determines the other two, for the known parameters of the far upstream fluid (Ferrand 2010, Pavlovi\'c 2018).

Finally, the downstream temperature is required in order to calculate $p_\mathrm{inj}$. By using ideal fluid equation of state $P\propto \rho T$ and Eq. (\ref{eq:bif99}), in the case of protons
\begin{equation}
\frac{T^\mathrm{p}_1}{T_0} = \frac{\rho_0}{\rho_1} \frac{P^\mathrm{p}_{\rm th,1}}{P_{\rm th,0}} = R_{\rm prec}^{\gamma-1} \left[1+\zeta(\gamma-1)\frac{M_{\rm S,0}^2}{M_{
\rm A,0}} (1-R_{\rm prec}^{-\gamma})\right],
\label{eq:T1}
\end{equation}
and
\begin{equation}
\frac{T^\mathrm{p}_2}{T^\mathrm{p}_1} = \frac{\rho_1}{\rho_2} \frac{P^\mathrm{p}_{\rm th,2}}{P^\mathrm{p}_{\rm th,1}}=\frac{(\gamma+1)R_{\rm sub}-(\gamma-1)\left[1-(R_{\rm sub}-1)^3\frac{P_{\rm w,1}}{P_{\rm th,p1}}\right]}{\left[(\gamma+1) - (\gamma-1)R_{\rm sub}\right]R_{\rm sub}} - \frac{\Delta E}{k T^\mathrm{p}_1}
\label{eq:T2}
\end{equation}
\textbf{is }obtained from Eq. (\ref{eq:vainio2}), except for an ad hoc introduced term containing $\Delta E$ that will be explained below. {Similarly, from Rankine-Hugoniout energy equation for electrons we have
\begin{equation}
\frac{T^\mathrm{e}_2}{T^\mathrm{e}_1} = \frac{\rho_1}{\rho_2} \frac{P^\mathrm{e}_{\rm th,2}}{P^\mathrm{e}_{\rm th,1}}= \frac{\gamma-1}{2}\frac{m_\mathrm{e}}{m_\mathrm{p}} M_{\rm S,0} ^2 (R_\mathrm{prec})^{-\gamma -1} \Big(1- \frac{1}{R_\mathrm{sub}^2}\Big) +1 + \frac{\Delta E}{k T^\mathrm{e}_1},
\label{eq:T2e}
\end{equation}
where ${T^\mathrm{e}_1}/{T_0} = R_{\rm prec}^{\gamma-1}$} The energy  $\Delta E \approx 0.3$ keV (see Ghavamian et al. 2007, 2013) is removed from Alfven-heated protons and added to electrons (constant electron heating ahead of the subshock), so that the downstream temperatures are $T^\mathrm{p}_2 = T^\mathrm{'p}_2 -  \Delta E /k$, $T^\mathrm{e}_2 = T^\mathrm{'e}_2 + \Delta E/k$ (where temperatures $T^\mathrm{'}_2$ are obtained from jump conditions).

We now search for the solution, by using the assumed $R_{\rm prec}$ and the initial conditions
\begin{equation}
U_{p} (p=p_{\rm inj}) = U_{p} (x=0^{-}) = \frac{u_1}{u_0} = \frac{1}{R_{\mathrm{prec}}},
\label{eq:diff1}
\end{equation}
\begin{equation}
\label{eq:limes}
\lim_{p\to p _{\rm{inj,p}}}f^\mathrm{p} _0 (p) =  \frac{3 R_{\rm{sub}}}{R_{\rm{sub}} - 1}
\frac{\eta n_{\rm{1}}}{4\pi p_{\rm{inj,p}}^3}.
\end{equation}
Between $p _{\rm{inj,e}}$ and $p _{\rm{inj,p}}$ we assumed $U_p = \frac{1}{R_{\mathrm{prec}}}$, so $f^\mathrm{e}_0 (p)\propto p^{-3R_{\rm{sub}}/(R_{\rm{sub}}-1)}$.
However, arbitrary chosen $R_{\rm prec}$ not necessarily satisfies the boundary condition
\begin{equation}
U_{p} (p=p_{\rm{max,p}}) = U_{p} (x\rightarrow -\infty) = \frac{u_0}{u_0} = 1,
\label{eq:diff2}
\end{equation}
which is used to end the integration at $p_{\mathrm{max,p}}$, so the solution is found iteratively.
To make the equations look more simple, we introduce the change in the variables:
$$
\frac{p}{m_\mathrm{p} c} \rightarrow p^*,
$$
$$
\frac{4\pi}{3} \frac{m_\mathrm{p}^4 c^5}{\rho _0 u_0^2} f_0 \rightarrow f_0^*.
$$
After this change, the system of non-linear equations (\ref{eq:step1}) and (\ref{eq:diff_cont}) that we solve numerically, takes the form:
\begin{equation}
\frac{1}{3}\Big( \frac{1}{R_{\mathrm{tot}}} - U_{p*}\Big) p^* \frac{\mathrm{d} f^\mathrm{*,p}_\mathrm{0}}{\mathrm{d} p^*} - \Big( U_{p*} + \frac{1}{3} p^* \frac{\mathrm{d} U_{p*}}{\mathrm{d} p^*} \Big) f^\mathrm{*,p}_\mathrm{0}= 0 ,
\end{equation}
\begin{equation}
\frac{1}{3}\Big( \frac{1}{R_{\mathrm{tot}}} - U_{p*}\Big) p^* \frac{\mathrm{d} f^\mathrm{*,e}_\mathrm{0}}{\mathrm{d} p^*} - \Big( U_{p*} + \frac{1}{3} p^* \frac{\mathrm{d} U_{p*}}{\mathrm{d} p^*} \Big) f^\mathrm{*,e}_\mathrm{0}= 0 ,
\end{equation}
\begin{eqnarray}
\frac{\mathrm{d}U_{p*}}{\mathrm{d}p^*}\left[1 - \frac{U_{p*}^{-(\gamma+1)}}{ M_{\mathrm{S},0}^2}\left(2 + \zeta(\gamma -1)\frac{M_{\mathrm{S},0}^2}{M_{\mathrm{A},0}}\right) - \frac{1-\zeta}{8 M_{\mathrm{A},0}} \frac{U_{p*}^2+3}{U_{p*}^{5/2}}\right] \nonumber \\
=  \frac{{p^*}^4 f^\mathrm{*,p}_0(p^*)}{\sqrt{1 + {p^*}^2}} + \frac{{p^*}^4 f^\mathrm{*,e}_0(p^*)}{\sqrt{(m_\mathrm{e}/m_\mathrm{p})^2  + {p^*}^2 }}.
\end{eqnarray}
The proton and electron advection-diffusion equations (37) and (38) are solved simultaneously in the iteration cycles, together with Eq.~(39).

\section{\label{sec:results}Results and Discussion}

\begin{figure*}[h!]
\centerline{\includegraphics[width=0.5\textwidth,keepaspectratio]{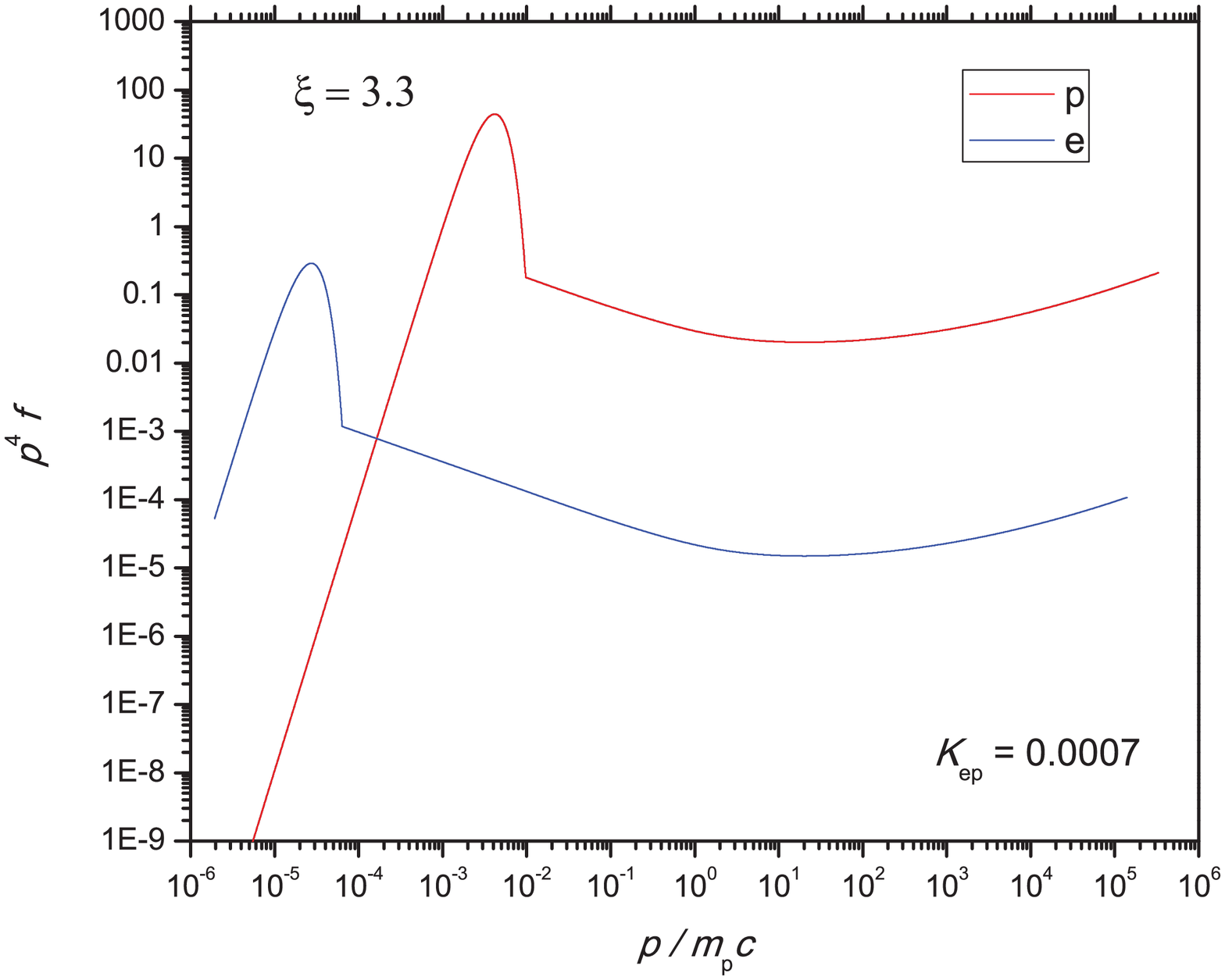}
\includegraphics[width=0.5\textwidth,keepaspectratio]{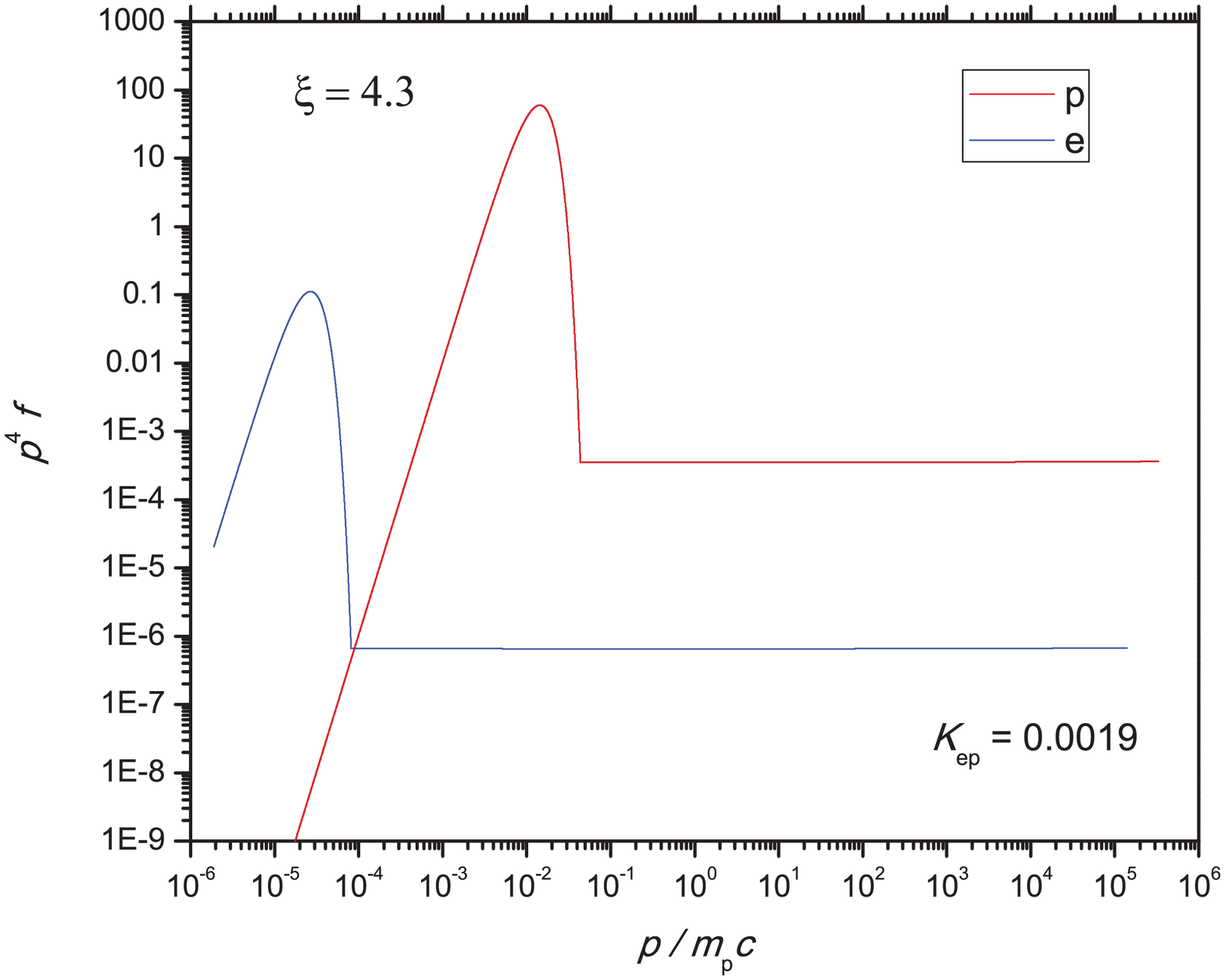}}
\begin{center}
\caption{\label{fig:nldsa_spectra} The proton and electron spectra for the injection parameter $\xi$ = 3.3 (left) and $\xi$ = 4.3 (right) in the case of realistic shock.}
\end{center}
\end{figure*}

\begin{figure}[h!]
\centerline{\includegraphics[width=0.6\textwidth,keepaspectratio]{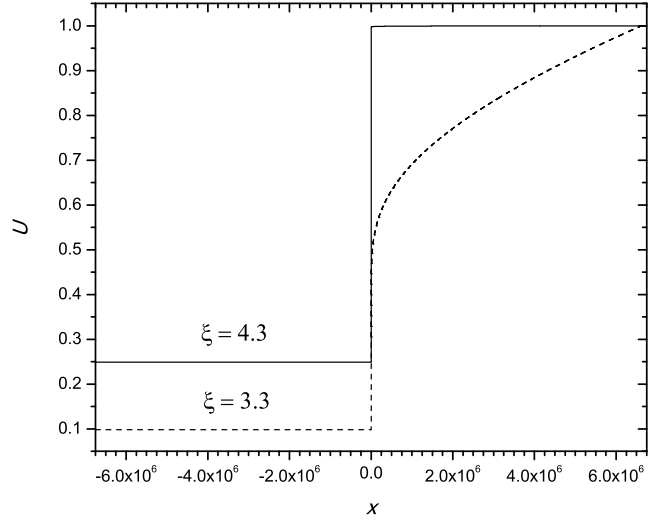}}
\begin{center}
\caption{\label{fig:graphux} The flow profiles $U=u/u_0$ for models with $\xi =3.3$ and $\xi =4.3$, assuming Bohm diffusion. Subshock is at position $x=0$ {($x>0$ upstream, $x<0$ downstream)}, $x$ being in units $m_p c/(q B_0)$ where $q$ is elementary charge and $B_0$ magnetic field.}
\end{center}
\end{figure}

From PIC simulations, we found $\eta_e \sim \eta_i$ (which implies $\xi_e \sim \xi_i$ in a non-relativistic case). We applied this rule to the model of non-linear DSA (that also includes constant electron heating) and calculated particle spectra (shown in Fig.~\ref{fig:nldsa_spectra}) for the two cases: strongly modified shock ($\xi = 3.3$), and unmodified shock (test particle case) with $\xi = 4.3$, as in Caprioli, Amato \& Blasi (2010). Otherwise, the parameters in both cases are the same and given for the realistic case: shock velocity is $u_0 = 5000\ \mathrm{km}\ \mathrm{s}^{-1}$, ambient density $n_\mathrm{H} \sim 0.1\ \mathrm{cm}^{-3}$, temperature $T_0 = 10^5$\ K, magnetic field $B_0 = 5.3775\  \mu \mathrm{G}$, sonic and Alfv\`enic Mach numbers are equal $M_{\mathrm{S},0} = M_{\mathrm{A},0} = 135$, and Alfven-heating parameter $\zeta  = 0.5$. The plots show thermal (Maxwellian) and non-thermal distributions that join at $p_\mathrm{inj}$. In the case $\xi =3.3$, the subshock and total compressions are $R_\mathrm{sub} = 3.093$, $R_\mathrm{tot} = 10.210$, with $K_\mathrm{ep} = 0.0007$; while for $\xi =4.3$, $R_\mathrm{sub} = 3.999$, $R_\mathrm{tot} = 4.018$, and $K_\mathrm{ep} = 0.0019$. In Fig.~\ref{fig:graphux}, we give the flow profiles in these two cases.

\begin{figure}[h!]
\centerline{\includegraphics[width=0.5\textwidth,keepaspectratio]{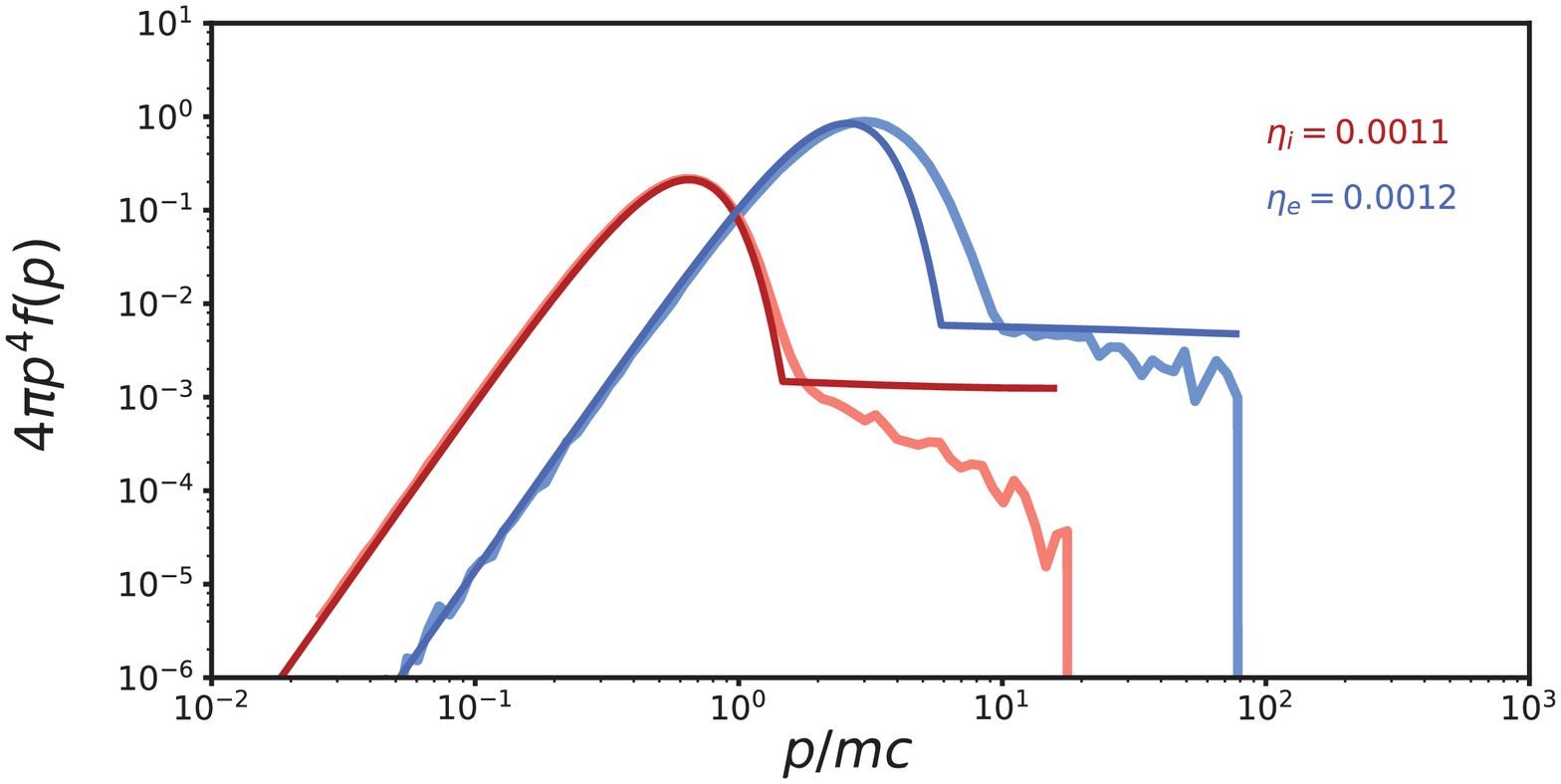}
\includegraphics[width=0.5\textwidth,keepaspectratio]{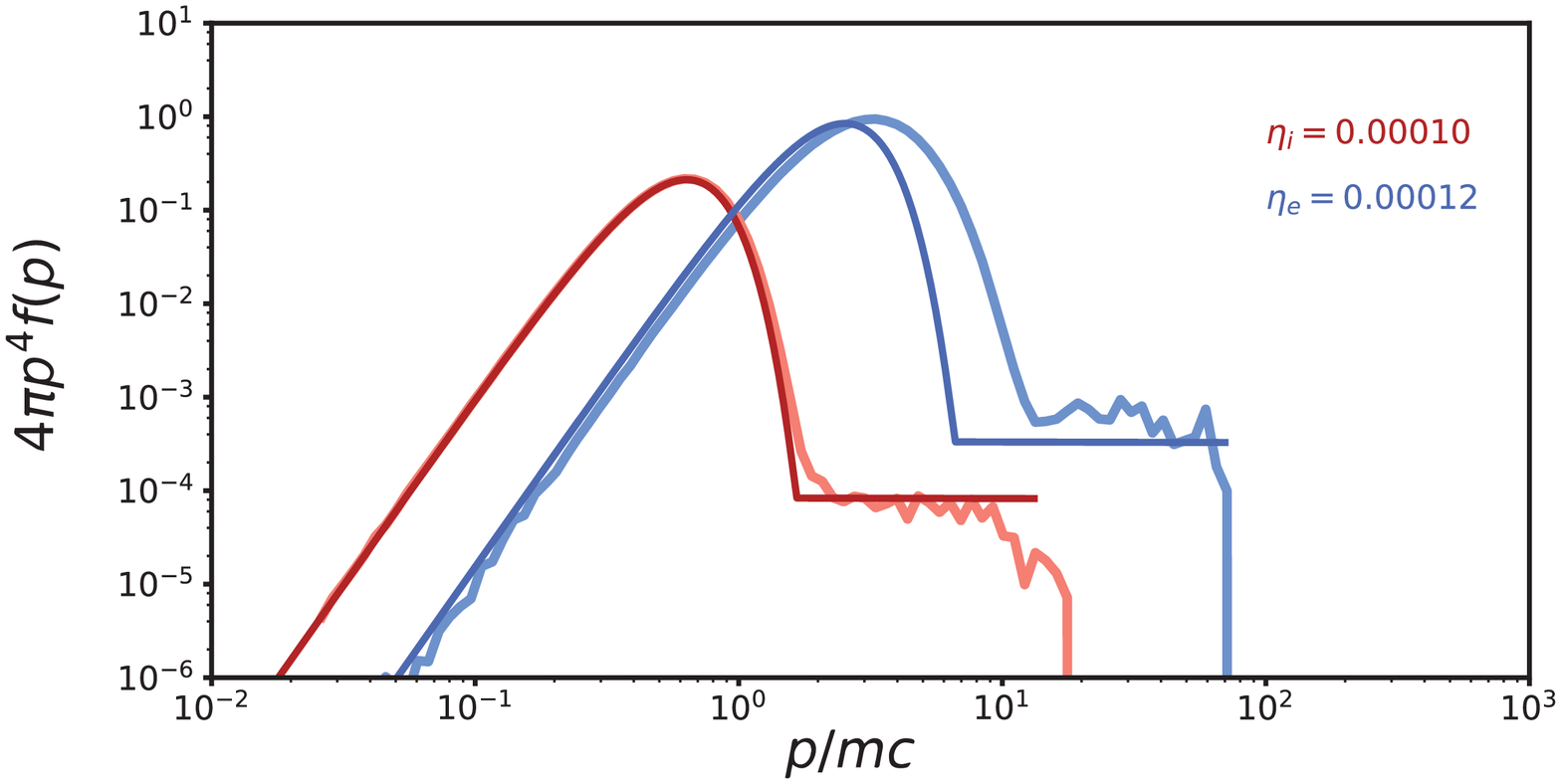}}
\begin{center}
\caption{\label{fig:compare_spectra} The ion (red) and electron (blue) particle spectra at the end of a PIC run 4, compared to the spectra calculated by our semi-analytical model for the non-relativistic case with the same Alfv\`en and sonic Mach numbers as in run 4. The top plot corresponds to the spectra in the near downstream where we measure $\xi \approx 3.2$, and the bottom plot corresponds to the far downstream where $\xi \approx 3.7$. On the horizontal axis, $p/p_{\rm sh}$ is given (where $p_{\rm sh} = \gamma_{\rm rel} m v_{\rm sh}$). On the verical axis, the normalized $4 \pi p^4 f(p)$ is given (where $p$ is in the units of $mc$).}
\end{center}
\end{figure}

\begin{figure}[h!]
\centerline{\includegraphics[width=0.6\textwidth,keepaspectratio]{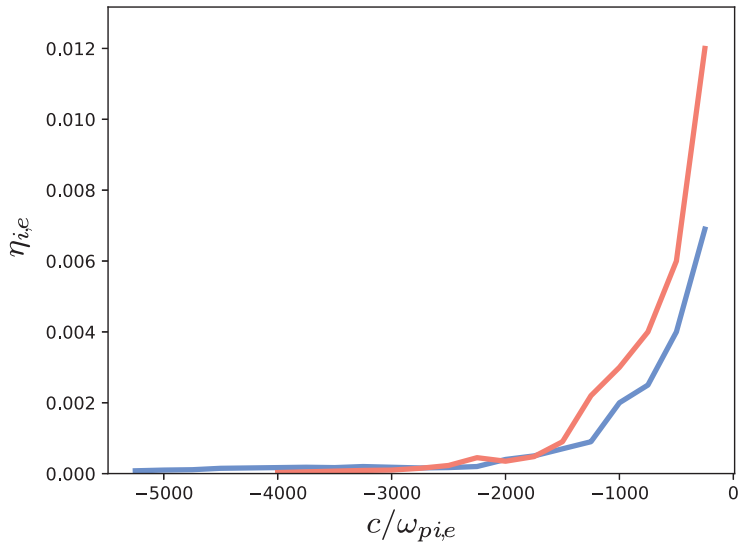}}
\begin{center}
\caption{\label{fig:eta-x} The amounts of CR ions $\eta_i$ (red line) and electrons $\eta_e$ (blue line) in the downstream, as functions of the distance from the shock. On horizontal axis, $x$-coordinate is given in the units of $c/\omega_{pi}$ and $c/\omega_{pe}$ for ions and electrons, respectively. The shock is at zero.}
\end{center}
\end{figure}

At the end of our long PIC run, we measure $\eta_i \sim 0.0001$ and $\xi_i \sim 3.7$, which matches the case of a weakly modified shock. The shape of the particle spectra in the near downstream (the upper graph in Fig.~\ref{fig:compare_spectra}) is more similar to the modified case with $\xi \sim 3.2$ and $\eta \sim 0.001$. However, it is interesting that this is a very local modification, and only a transient, which leads to the difference between the measured and calculated ion and electron spectra (only in the slopes, while $\eta_i \sim \eta_e$) in Fig.~\ref{fig:compare_spectra}. Farther in the downstream, the spectra flattens to $f(p) \sim p^{-4.2}$, with $\xi \sim 3.7$ and $\eta \sim 0.0001$, which implies the very weak shock modification. As ions had enough time to accelerate, and thus, to populate the non-thermal tail, the model spectra and the PIC spectra coincide in the far downstream (bottom graph). Also, as electrons evolve faster than ions, the ion spectra that is in the same stage of evolution corresponds to the region which resides deeper in the downstream, at the distance $\sim \sqrt{m_i/m_e}$ times farther relative to electron spectra. In Fig.~\ref{fig:eta-x}, we show how $\eta_{i,e}$ changes in the downstream. Once the distance from the shock is scaled to the particle skin depth, the ion and electron $\eta$-profiles become similar. The calculated electron spectra in Fig.~\ref{fig:compare_spectra} do not match the measured electron spectra, because it is given in a non-relativistic case. As the ion Maxwellian resides in the region of velocities less than the speed of light, the calculated and measured ion spectra coincide. The density compression in PIC runs is slightly above $\sim 4$ in the whole downstream region, which implies that there is almost no modification to the shock (although, by the near downstream spectra, the shock looks like being locally modified).

From the non-linear model, we find $K_\mathrm{ep}$ as a function of the Mach number. Electron-to-proton ratio at high energies, assuming $\eta _e = \eta _p$ (i.e. $\xi _e = \xi _p$ in a non-relativistic case), is
\begin{equation}
K_{\mathrm{ep}} = \frac{\eta _e}{\eta _p} \left( \frac{p_\mathrm{inj,e}}{p_\mathrm{inj,p}}\right)^{\frac{3}{R_{\mathrm{sub}}-1}} =
{\left(\frac{m_\mathrm{e} T_2^\mathrm{e}}{m_\mathrm{p} T_2^\mathrm{p}}\right)}^{\frac{3}{2(R_{\mathrm{sub}}-1)}} = \left(\frac{m_\mathrm{e}}{m_\mathrm{p}} \beta \right)^{\alpha},
\end{equation}

\begin{equation}
\beta \approx \frac{\frac{2(\gamma -1)}{(\gamma +1)^2} m_\mathrm{e} u_0^2 + \Delta E}{\frac{2(\gamma -1)}{(\gamma +1)^2} m_\mathrm{p} u_0^2},
\end{equation}

\noindent where $\alpha$ is the so-called spectral index, and $\beta$  is the downstream temperature ratio, which is given here for strong shocks ($M_{\mathrm{S},0} \rightarrow \infty$, Ghavamian et al. 2013) with electron heating (the full expression can be obtained from Eqs. (\ref{eq:T2}) and (\ref{eq:T2e})).

\begin{figure*}[h!]
\centerline{\includegraphics[width=0.49\textwidth,keepaspectratio]{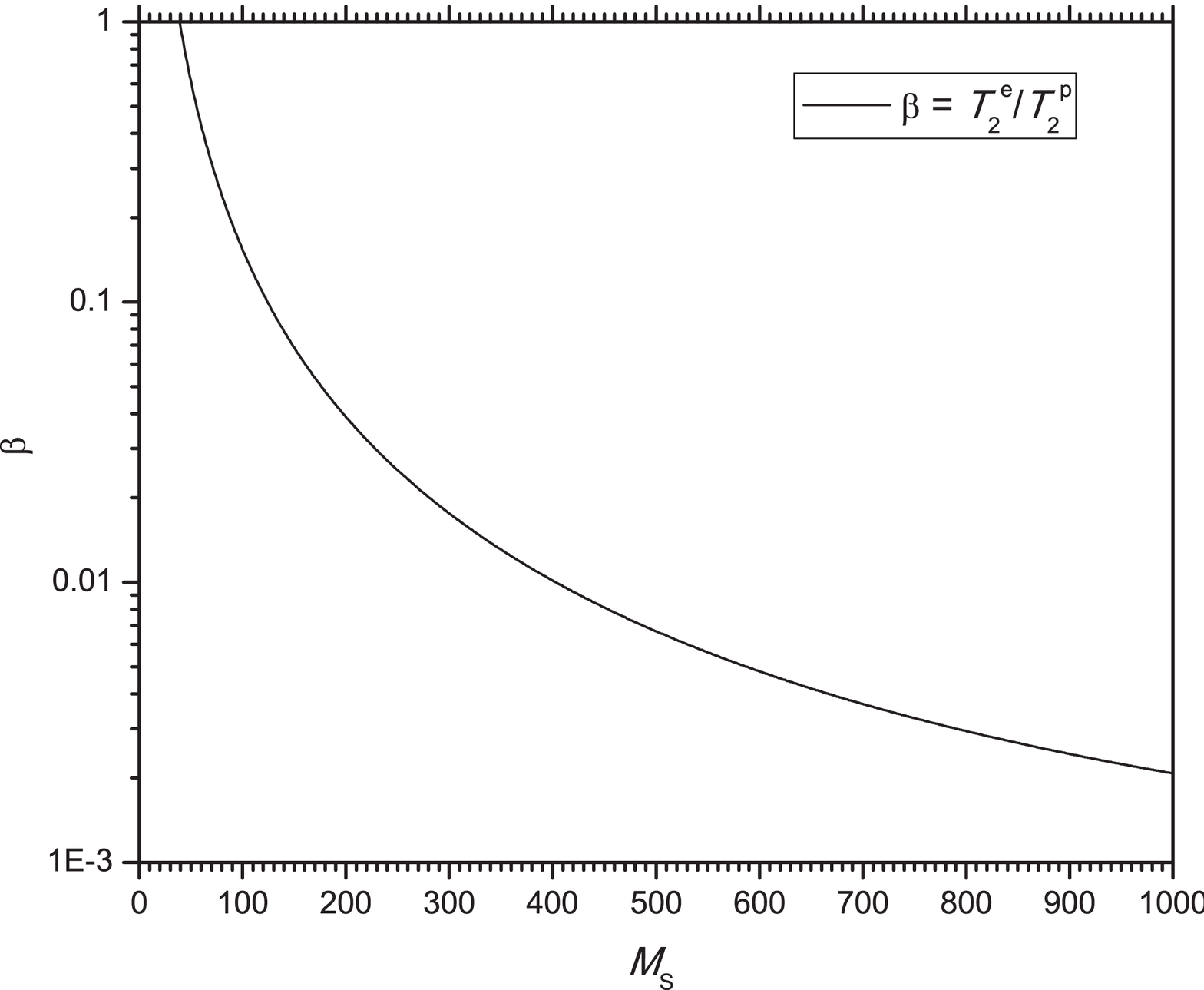}
\includegraphics[width=0.5\textwidth,keepaspectratio]{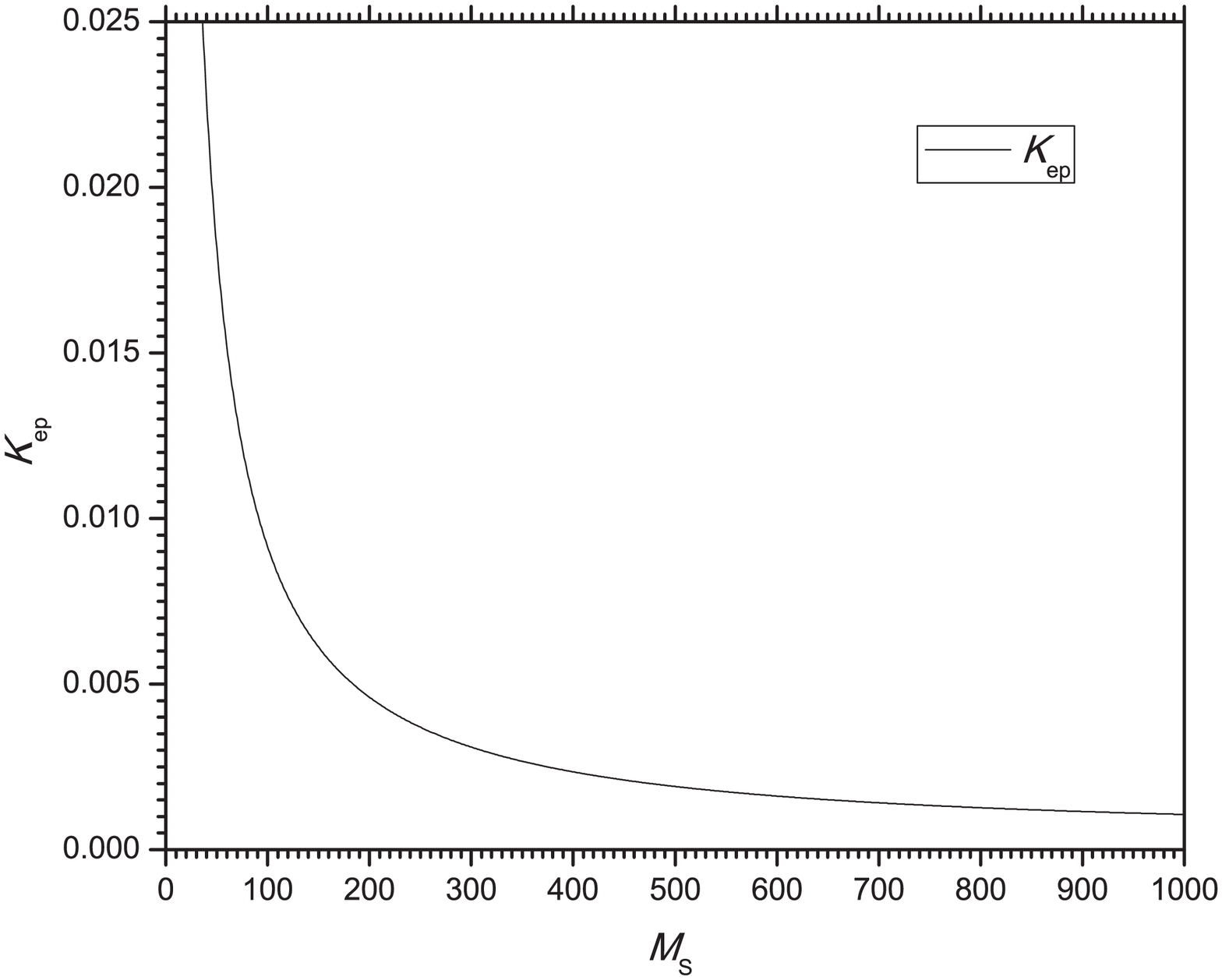}}
\begin{center}
\caption{\label{fig:ratios} Downstream temperature ratio $\beta$ (left) and electron-to-proton ratio $K_\mathrm{ep}$ (right), as functions of the sonic Mach number {$M_{\mathrm{S}}$ calculated for an assumed total} ambient sound speed of 10 km s$^{-1}$.}
\end{center}
\end{figure*}

In Fig.~\ref{fig:ratios}, we plot a more general $\beta$-law (for any {$M_{\mathrm{S}}$,} still including $\Delta E$, but in the test-particle regime) for an assumed {total} ambient sound speed of 10 km s$^{-1}$, and we give an analytical approximation to $K_\mathrm{ep}$ with such $\beta$.  We find that for this modified Rankine-Hugoniot shock jump conditions in the test particle regime, the observed ratio for Galactic cosmic rays $K_\mathrm{ep} \sim $ 1:100 corresponds to the Mach number $\sim 100$  (shock velocitiy $u_0 \sim$ 1000 km s$^{-1}$). For modified shocks, the $K_\mathrm{ep}$ is not a simple function of the Mach number (as in the unmodified case shown in Fig.~\ref{fig:ratios}), but it depends in a rather complex way on model input parameters (such as $\xi$, $\zeta$, $M_{\mathrm{S},0}$, and $M_{\mathrm{A},0}$). However, even this simplified model agrees well with multi-wavelength observations of young SNRs suggesting that $K_\mathrm{ep} \sim 10^{-3}$ or less (V\"olk et al. 2005, Morlino \& Caprioli 2012).

We assume that the decrease in parameter $\beta$ (and thus in $K_\mathrm{ep}$) is significantly slower in our non-linear DSA model, due to the  applied constant electron heating of $\Delta E = 0.3~\rm keV$ (which is found from observations; Ghavamian et al. 2007, 2013), than in the case without heating. In order to match our semi-analytical model to PIC run 4 simulation, we also needed to apply electron heating of $\sim$ 150 eV for $\xi = 3.2$ case and $\sim 208$ eV for $\xi = 3.7$ case.

\section{\label{sec:conclusions}Conclusions}

The model we presented in this paper is quite different from all the recent models which use the assumptions that ions and electrons are injected into DSA with the same momentum or energy. Instead, we introduced the novelty in the injection conditions ($\eta_i \sim \eta_e$, $\xi_i \sim \xi_e$) that we found from the self-consistent PIC simulations. Even though in the cases given in  Figs. 8 and 11 we assumed $\eta _e = \eta _p$ (that seems reasonable to us), one can easily drop this assumption in the injection recipe, and apply Eq. (4) separately for protons and electrons, which will consequently also affect $K_\mathrm{ep}$, but basically not change our prescription that the injection of electrons into DSA can be treated in a somewhat analog manner to protons. We applied this injection recipe to a model of non-linear DSA with the constant electron heating included, and we obtained the spectra in some final, quasi-stationary stage of the shock evolution. Whilst looking at the larger scale, the shocks in our PIC runs correspond only to some very first stages of the non-linear modification, showing the weak precursor in the upstream. We found that $\eta$ is varying in the downstream, implying the transient modification right behind the shock (with $\xi \sim 3.2$ and $\eta \sim 0.001$), and only the weak modification in the far downstream (with $\xi \sim 3.7$ and $\eta \sim 0.0001$). Will the relation $\eta _e \sim \eta _p$ hold for any Mach number used in PIC simulations, will the shock eventually become modified there, and what is the physics that will lead to such modification, are the issues that we plan to address in our forthcoming work.

This work is the first such attempt to calculate the ion and electron spectra in consistency with kinetic simulations and, thus, to overcome the gap between the micro and macro-physics. It can be of a great importance in both, practical (observational) and theoretical aspects. It is our belief that this model has a potential in explaining the overall abundances of cosmic ray electrons and ions. This model, however, does not take into account the effects of CR transport in the Galaxy, which are shown to be important for protons and heavier nuclei (Evoli et al. 2019). Also, the transport of leptons is loss dominated down to energies of the order of tens of GeV (Evoli et al. 2020). According to the simplified $K_\mathrm{ep}$ dependence presented, the observed electron-to-proton ratio for Galactic cosmic rays 1:100 could majorly originate from the shocks with the velocity $\sim$ 1000 km s$^{-1}$. The model can definitely find its application in the calculation of particle spectra of SNRs and related objects, and in the modeling of electron synchrotron emission from these sources, allowing us to gain knowledge of physical parameters of ISM shocks.  

Although our model well reproduces the spectra that we get from PIC simulations, it needs additional verification for self-consistency in the case of a strongly modified shock. For a shock to reach significant level of modification, PIC simulations need to be pushed much further. Also, it is shown by Diesing \& Caprioli (2019) that the electron synchrotron losses are significant and should be considered in calculating the electron spectra. The electron heating parameter needs to be tested for a dependence on Alfv\`enic Mach number and eventually constrained by the results of PIC simulations. We leave such a closure to be conducted in the succeeding work.

\section*{Acknowledgements} We thank Anatoly Spitkovsky and both Reviewers for commenting and pointing to some important issues in the preliminary version of the paper. The PIC simulations were run on the PARADOX-IV supercomputing facility at the Scientific Computing Laboratory of the Institute of Physics Belgrade, on cluster JASON of Automated Reasoning Group (ARGO) at the Department of Computer Science, and on a new cluster SUPERAST at the Department of Astronomy, Faculty of Mathematics, University of Belgrade. The results of PIC simulations were in part visualized by ISEULT - a GUI written by Patrick Crumley. The authors acknowledge the financial support of the Ministry of Education, Science and Technological Development of the Republic of Serbia through the contract No. 451-03-68/2020-14/200104.

\end{document}